
\magnification=\magstep1
\baselineskip=20 pt plus 1pt minus 1pt
\vsize=8.9truein
\hsize=6.8truein
\hfuzz=1.5pt
\footline={\ifnum\pageno=1 \hfil\else\centerline{\folio}\fi}
\def\v{\vskip6pt}
\def\veject{\vfill\eject}
\def\hi{\hangindent=15pt}
\def\v{\vskip 6truept}
\def\vv{\v\v}
\def\veject{\vfill\eject}
\def\no{\noindent}
\def\h{{\textstyle {1 \over 2}}}
\def\p{\partial}
\def\setify#1{{$\{#1\}$}}
\def\qcomma{\quad,\quad}
\def\qqcomma{\qquad,\qquad}
\def\E#1{Eq.~(#1)}
\def\Es#1{Eqs.~(#1)}
\def\complx{{\fam=11 \cmpl}}
\def\zm{z^\mu}
\def\sm{^\mu}

\def\bth{{\overline\theta}}
\def\ee{\hbox{\bf{e}}}
\def\cl#1{{{\cal{#1}}}}
\def\mm{{\bc{{\cl{M}}}{\ell}}}

\def\pp{{\cl{B}}}
\def\aa{{\cl{A}}}

\def\Four{\int_{-\pi}^{\pi}\!\!}

\def\fc#1#2{\mathop{{#1}}\limits_{\raise2pt\hbox{\smash{\hbox{$
	\scriptscriptstyle{\!#2}$}}}}}

\def\ad#1{{(\hbox{ad}\,\b #1)}}
\def\b#1{{\bf \uppercase{#1}}}
\def\ww#1{{{\vec{\underline{#1}}}}}
\def\w#1{{\underline{#1}}}
\def\ad#1{{\rm ad\,#1}}
\def\cp{{\cal{P}}}
\def\cq{{\cal{Q}}}
\def\ck{{\cal{K}}}
\def\pse{{\Phi_{{}_t}{}_{\scriptscriptstyle(P)}}}
\def\ft#1{${}^{#1}$~}
\def\bc#1#2{\mathop{{\bf{#1}}}\limits_{\raise2pt\hbox{\smash{\hbox{$
 	\scriptscriptstyle{\!#2}$}}}}}

\font\chapter=cmbx12 scaled \magstep 1
\font\smchapt=cmbx12
\font\abstrct=cmr8
\font\reglar=cmr10
\font\sans=cmssbx10 scaled \magstep 1
\font\hscript=eufb10 scaled \magstep 1
\font\outline=msbm10

\font\cyril=wncyr10 scaled \magstep 1
\font\trila=cmmib10
\font\euss=eusb10 scaled\magstep 1
\textfont9=\trila
\textfont10=\cyril
\textfont11=\outline
\textfont12=\hscript
\textfont13=\sans
\textfont14=\euss
\mathchardef\alpha="710B
\mathchardef\beta="710C
\mathchardef\gamma="710D
\mathchardef\delta="710E
\mathchardef\theta="7112
\mathchardef\lambda="7115
\mathchardef\cmpl="7143
\mathchardef\russ="715A
\mathchardef\gotha="7141
\mathchardef\gothb="7142
\mathchardef\gothk="714B
\mathchardef\calef="7146
\mathchardef\caleg="7147

\def\oua{{\fam=11 A}}   
\def\oub{{\fam=11 B}}   
\def\real{{\fam=11 R}}
\def\qa{{\fam=12 A}}
\def\qb{{\fam=12 B}}
\def\sna{{\fam=13 A}}
\def\snb{{\fam=13 B}}
\def\snm{{\fam=13 M}}
\smchapt
\centerline{Infinite-Dimensional Estabrook-Wahlquist Prolongations}
\centerline{for the sine-Gordon Equation}
\reglar\vv\baselineskip=14pt
\centerline{J. D. Finley, III and John K. McIver}
\v\baselineskip=11pt
\centerline{Department of Physics and Astronomy}
\vskip3truept
\centerline{University of New Mexico}
\vskip3truept
\centerline{Albuquerque, N.M.  87131}
\vv
{\abstrct \baselineskip=10pt
We are looking for the universal covering algebra for all symmetries of a given
pde, using the sine-Gordon equation as a typical example for a
non-evolution equation.  For non-evolution equations, Estabrook-Wahlquist
prolongation structures for non-local symmetries depend on
the choice of a specific sub-ideal, of the contact
module, to define the pde.  For each inequivalent such choice we determine the
most
general solution of the prolongation equations, as sub-algebras of the
(infinite-dimensional) algebra of all vector fields over the space of non-local
variables associated with the pde, in the style of Vinogradov covering spaces.
We show explicitly how previously-known prolongation structures, known to lie
within the Kac-Moody algebra, $A_1^{(1)}$, are special cases of these general
solutions, although we are unable to identify the most general solutions with
previously-studied algebras.
\par We show the existence of gauge transformations between prolongation
structures, viewed as determining connections over the solution space, and
use these to relate (otherwise) distinct algebras.  Faithful realizations of
the universal algebra allow integral representations of the prolongation
structure, opening up
interesting connections with algebras of Toeplitz operators over Banach spaces,
an area that has only begun to be explored.\medskip}
\reglar
\baselineskip=20pt plus 2pt minus 2pt
\no {\bf I.  Universal Algebras for Symmetries}
\v
The sine-Gordon equation has been applied to physically interesting problems
for over a hundred years;\ft{1} quite a large list of applications
is given by Rogers and
Shadwick.\ft 2.  The generalized form $u_{xy} = f(u)$ has been the subject of
numerous investigations into questions of symmetries.\ft{3-9}
Geometrical questions concerning it occasioned the
creation of the first B\"acklund transformation,\ft{1,10} which we view
as involving a non-local symmetry, since it relates
the dependent variables in two distinct copies of the original equation.
A particular approach to determining non-local symmetries was created by
Estabrook and Wahlquist,\ft{11} based on Cartan's\ft{12} method of describing
pde's via the use of differential forms.  They refer to the new variables
allowed by
their non-local symmetries as {\it pseudopotentials}, and the associated
algebras as
{\it prolongation structures}.  This method is quite
popular\ft{13-17}, and has, of course, been used to study our form of the
 sine-Gordon equation.\ft{18-22}
Since we believe these prolongation structures
are an important tool\ft{23} in the general classification
of ``equivalent'' pde's, or systems of pde's, and
in the general determination of their solutions,
we, along with many others,\ft{24-29} are quite interested in the determination
of the most general such
structures, and have found the sine-Gordon equation useful in one aspect of
this.
\v
A k-th order pde may be realized as a subvariety, $Y$, of a finite jet bundle,
$J^{(k)}(M,N)$, where $M$ is intended as
the space of independent variables and $N$ the
space of dependent variables in the original
partial differential equation.
The search for
generalized symmetries of a pde is most easily performed on the
infinite prolongation of that pde, to
$Y_\infty \subset J^\infty$, as described, for instance, by
Vinogradov.\ft{30}, while the further prolongation to {\it coverings} of
$J^\infty$, by Vinogradov and Krasil'shchik (VK),\ft{31-32}
allows the determination of non-local symmetries as well.
Beginning with $\p_a\equiv \p/\p x^a$
as a choice of basis for tangent vectors over the base manifold, $M$,
the total derivative operators, $D_a$, are the (standard) lifts of these
vector fields to the infinite jet bundle, which still commute:
$[D_a\,,\,D_b\,] = 0$.  The covering space being searched for will
have fibers, $W$ over $J^\infty$, where a choice of
coordinates, \setify{w^A}, may be thought of as
allowed pseudopotentials, or
non-local variables, for the pde in question.
The further prolongation of the total derivative
operators into these fibers is accomplished by the addition of some
vector fields vertical with respect
to the fibers, i.e., $\b X_a = \sum X_a^A(\p/\p w^A)$,
where the coefficients depend on the jet variables as well as
the \setify{w^A} themselves.
In general these vector fields, $D_a + \b X_a$, would no longer commute
with each other; however, the requirement that they commute when
restricted to the subspace defined
by the original pde is exactly the requirement needed in order to ensure
that these additional
coordinates can in fact act as pseudopotentials for that pde:\ft{31}
$$\eqalign{0\ = \
\left[D_a+\b X_a\,,\,D_b+\b X_b\,\right]_{\big|_{Y_\infty
\times W}}\!\!\!\!& \equiv\left[\overline
D_a + \b X_a\,,\,\overline D_b + \b X_b\,\right]\cr & =
\left\{\overline D_a(X_b^C) - \overline
D_b(X_a^C)\right\} {\p\over \p w^C}\ +\ \left[\b X_a\,,\,\b X_b\,
\right]\quad,\cr} \eqno(1.1)$$
\no where the overbar on the total derivative operators indicates
that they have been restricted to
$Y_\infty$.   The general solution of these equations for the $\b X_a$
will describe all
possible coverings associated with this pde.
This requirement is of course reminiscent of the usual sort of
``zero-curvature'' requirements,\ft{33,34}  the distinctions
being that we specify first the pde, rather than the
particular algebra within which the
$\b X_a$'s reside, {\bf and} that we do not specify in advance the dimension of
the covering fibers.
\v As this equation is an identity in the coordinates of the jet
bundle, several independent equations will be determined.
Some of these equations will explicate some (or all) of the
dependence on the jet variables of the $\b X_a$, but none of their
dependence on the fiber variables, \setify{w^A}, so that each $\b X_a$
is now given in terms of linear combinations of vector
fields $\b W_\alpha$ with coefficients that depend only on the
\setify{w^A}.  The constraints on the
$w^A$-dependence is then encoded in
the commutators of these \setify{\b W_\alpha} among themselves.  The
solution to \E{1.1} will already have made specific requirements
concerning the values for {\bf some}
of the commutators of the \setify{\b W_\alpha}. Since these vector fields lie
within the entire
algebra of vector fields over the fiber, $W$, the universal algebra in which we
are interested is
the smallest subalgebra that maintains the linear independence of all of the
\setify{\b W_\alpha} and
faithfully reproduces the values of those of their commutators required by
\E{1.1}.  This subalgebra
will usually be infinite-dimensional.  It
determines the general solution to the covering problem, and all faithful
realizations of it, by making particular choices for the number of
pseudopotentials. We believe
this algebra is a universal object for the given pde and other pde's related to
it via some (local
or nonlocal) symmetry transformation; i.e., it may be used to characterize
related classes of
pde's.\ft{23,24,32}   The isolation and identification of such algebras is an
important part of the
process of determining and understanding all the solutions of nonlinear pde's.
Vector-field
realizations of this algebra can be used to generate B\"acklund
transformations, inverse scattering
problems,\ft{5,24} etc., although such complete
realizations of the entire algebra are not usually
necessary to obtain these transformations.  Infinite-dimensional algebras were
originally not
considered a profitable direction for study, because of the difficulty with
their
identification.  In their work on the generalized sine-Gordon equation,
Dodd and Gibbon\ft{22} already noted that they were left with an
infinite-dimensional
algebra; however, since they could map it
homomorphically to ${\bf sl}(2,\complx)$,
they did not ``need'' to consider the entire algebra.
Beginning with the work by van Eck,\ft{25} and Estabrook,\ft{28} on
identification of these
universal algebras for the KdV equation, such an approach was extended
considerably
by Hoenselaers and co-workers\ft{17,20,24}, by Omote\ft{19}, and by the group
at
Twente, who seem to have made this a studied art-form.\ft{15,26}
\v Two general statements have yet to be
made concerning the determination of such algebras.
Firstly, the dependence of the $\b X_a$ on the
various jet variables is determined by algebraic equations
if the original pde is an evolution
equation, i.e., of only first order with respect to one of the variables.
On the other hand, for
equations of higher order, the determining equations are (linear) partial
differential
equations, thus substantially increasing the
difficulty of finding the general solutions.  In the most general cases, these
equations have not been solved; nonetheless, a small set of additional
assumptions concerning the dependencies on the jet variables, will in
fact allow general solutions, as we intend to demonstrate.
\v The second distinct property associated with non-evolution equations
is that they allow distinct {\it subideals} of the entire ideal of total
derivative operators to remain {\it effective}\ft{2,13} as complete
descriptions
of the original pde.  That is, each such subideal retains sufficient
information to completely re-construct the pde; the existence of such
subideals correlates with the
existence of characteristic vectors for the pde, and, at least when there
are only two independent variables, requires that the pde be higher than
first order in both variables.  These questions were already discussed
by Pirani and Shadwick,\ft{13,27} who showed that the existence
of various, distinct subideals of the contact module
causes different prolongation algebras to be generated.
It is exactly this situation which shows up the significant differences
between the methods of EW, relative to those of
VK.  As described above, VK use the entire, restricted ideal of total
derivative
operators, $\overline D_a$, while the essential essence of
the EW approach, via differential forms, is to use a (differentially closed)
subideal of the entire contact module.  The smaller size of this
subideal reduces the generality allowed to
the $\b X_a$, and therefore reduces the number of independent vector fields,
\setify{\b W_\alpha}, {\it without losing any generality in the
pde.}
\v  A major purpose of this article is to
explicate more fully how these differences can lead to major differences in
the prolongation algebra generated, and to outline how those algebras are
related.  In addition, we will show that distinct choices of the
coordinates in the covering fibers, $W$, can obscure the differences
in these algebras, especially when one does not investigate the maximal
algebras created by the different subideals.  These differences in
coordinate choices may be viewed as related by {\it gauge transformations}
between different solutions of \Es{1.1}.  This viewpoint
is not as easily seen in the EW approach, as it is
in the VK approach, via total derivatives, where the
$\b X_a$ may be thought of as connections over the covering manifold,
and gauge transformations are a very natural thing.\ft{34, 35}
\v
We have found the generalized form of the sine-Gordon equation to be
an ideal vehicle for the understanding of these differences.
 It is true that one might easily think that
all possible interesting questions had been already been
answered for this equation.
Nonetheless, research on new properties of this equation, and new ways of
looking at it, seem to still be surfacing today.\ft{36}
In our case, it is actually the detailed knowledge of the properties of this
equation and its solutions that allows us to use it to describe and discuss
these previously-unanswered questions concerning the
meaning, behavior and use of the EW procedure.
Therefore, we will consider the equation,
$$ u_{xy} \quad = \quad f(u)\quad,\eqno(1.2)$$
where we do {\bf not} allow the function $f$ to depend on the independent
variables, \setify{x,y}.
(There are in fact some interesting aspects that occur when such
dependence is allowed; a separate report on a typical and important equation of
that type is being made.\ft{37}) We will consider all the non-isomorphic ideals
of 2-forms that can
be used to generate this equation, and will in each case describe the
general solution
to the determining differential equations and detail those commutation
relations for the associated vector fields $\b W_\alpha$ that the original
pde requires.
We give sufficiently complete descriptions of these algebras to show
why the identification question is particularly difficult for them.
We also show how each of them has {\it homomorphisms} onto quite large
subalgebras which are gauge equivalent.\v
Having already described a $k$-th order pde as a submanifold, $Y \subset
J^k(M,N)$, a solution to the pde is then
locally a map $u:U\subseteq~M \rightarrow N$ such that, $\forall x
\in U$, the section $j^k_xu$ lies entirely within $Y$.
{}From this point of view, Cartan's\ft{12} approach to pde's may be said to
begin with the `contact module,'
 $\Omega^k(M,N)\subseteq [J^k(M,N)]_*$, which is used to
determine whether a given section of $J^k(M,N)$ is the lift of
a function, $u$, on the base manifold.  It is generated by the
following set of 1-forms:
$$\Omega^k (M,N): \left\{ \matrix{ \theta^\mu = \,d\zm - \zm_a d x^a ,\hfill
\cr\cr
\theta^\mu_a = \,d\zm_a - \zm_{ab} dx^b, \hfill \cr
\qquad\qquad\quad\ldots \hfill\cr
\theta^\mu_{a_1a_2 \ldots a_{k\!-\!1}} = \,d\zm_{a_1a_2 \ldots a_{k\!-\!1}} -\>
\zm_{a_1
a_2 \ldots a_{k\!-\!1} a_k} dx^{a_k}\hfill\quad, \cr } \right. \eqno(1.3)$$
\no where the summation convention has been used with respect to repeated (once
upper, once lower) indices, and a choice for a local coordinate chart for a
neighborhood in $J^k(M,N)$ is given by
$\left\{x^a,\zm, z\sm_a, z\sm_{a_1a_2},\ldots, z\sm_{a_1 \ldots a_k}
\right\}$.
(The $z\sm_{a_1 \ldots a_q}$ are symmetric in their subscripts.)
\v The contact module `remembers' the relation that
the various coordinates of the jet bundle are `supposed to have' when one is
dealing with an actual function, i.e., when a section has been chosen.
Therefore the contact module vanishes when pulled back to $M$ by a function
$u:U\subseteq M\longrightarrow N$, i.e.,
$(J^ku)^*(\Omega^k) = 0$.  The ideal, ${{\cal I}}$, is the
differential closure of the
pullback of the contact module to $Y$.  It constitutes
the Cartan description of the original pde, and is
completely equivalent (and dual) to the description via the variety
$Y$ and the restricted total derivative operators.
\v
The essence of the EW procedure, for 2 independent variables,
 is to first determine an effective, proper subideal, $\ck\subset
{{\cal I}}$, generated by a set of 2-forms over $J^{k-1}$, to describe
the pde of interest.   Such smaller ideals have long been used to search
for potentials for pde's, and are characterized in Refs. 2 and 16 as
``effective.''  EW then
append to this ideal a set of contact
1-forms for the to-be-determined pseudopotentials, $w^A$:
$$\omega^A = - dw^A + F^A dx + G^A dy, \quad A = 1, \ldots, N\qquad,
\eqno(1.4)$$
\no where we will, however, now begin using EW's
symbol $\b F$ for VK's $\b X_x$ and EW's symbol $\b G$ for VK's $\b X_y$, with
the $F^A$ and $G^A$ as labels for their respective coefficients.
\no In the more classical approach, such a potential would only depend
on the jet variables.  When one potential is found it may be appended to the
original variables, and the enlarged system again searched for potentials.
In principle this process generates a (possibly infinite) sequence of
potentials.  The original contribution of Estabrook and Wahlquist was to notice
that one could
simply suppose the existence of an as-yet-undiscovered space $W$---the
fibers of VK's covering space---of such
potentials and look for them all at once, in which case they were called
pseudopotentials.
\v The EW approach allows the $F^A$ and
$G^A$ to depend upon themselves; therefore proper phrasing of
 the closure question now requires that
the $d\omega^A$ be contained within the extended ideal, ${\cal
K}\oplus\{\omega^i\}$.  Labelling the
(2-form) generators of ${\cal K}$ by $\{ \alpha^r|\,r= 1, \ldots , p\}$, this
means we are searching for pseudopotentials $w^A$ that allow
the existence of some functions $f^A{}_r$ and 1-forms $\eta^A{}_B$
such that    $$dF^A \wedge dx + dG^A\wedge dy = f^A{}_r \alpha^r + \eta^A{}_B
\wedge \omega^B\qquad. \eqno(1.5)$$
\v Allowing dependence of $F^A$ and $G^B$ on all of the variables of $J^{k-1}
\times W $, and comparing coefficients of the various independent 2-forms
 on both sides of \E{1.5}, three sorts of information are acquired.  Firstly,
we determine on which jet variables they are
allowed to depend.  Secondly, we are able to relate the values of the
previously-unknown Lagrange multipliers,
$f^A{}_r$ and $\eta^A{}_B$, to various derivatives of the $F^A$ and the $G^B$.
      While the continuation of the prolongation process, toward eventually
finding
new solutions of the original pde, does not explicitly require that we
know the values of these Lagrange multipliers, it is nonetheless
quite important that the final choices of
$F^A$ and $G^B$ should be such as to maintain {\bf non-zero} the multipliers,
$f^A{}_r$.  After all they retain the information needed by the procedure to
``remember''
the original ideal, ${\cal {K}}$, i.e., to remember the pde with which the
process began.
Lastly, but importantly, the coefficients of $dx \wedge dy$ in \E{1.5}
generate a commutator equation for these vertical vector fields
which always has the following general form:
$$[{\b F} + \p_x, {\bf G} + \p_y] = \sum_\sigma a_\sigma{\p\over\p z_\sigma}{\b
F}\> +\>
\sum_\sigma b_\sigma{\p\over\p z_\sigma}{\bf G}\quad,\eqno(1.6)$$
\no where the sum is over all coordinates of $Y$ and the coefficients
$a_\sigma$ and $b_\sigma$ depend on the specific system being considered.
This equation is in fact identical to \E{1.1}, and makes explicit the
connection between the EW approach to the calculations and the VK
approach, that begins with zero-curvature equations similar to those
of Zakharov and Shabat.\ft{33}
\veject
\no {\bf II.  The Prolongation Problem for the (Generalized) sine-Gordon
Equation}
\v We now begin explicit applications of these procedures to our \E{1.2},
choosing coordinates on the solution space $
Y\subset J^2(M,N)$, by
eliminating all the ``mixed derivatives'' of $u$, so that ${{\cal I}}$, the
contact module restricted to $Y$, is generated by
$$\eqalign{\bth_u  \equiv &du -p\,dx -q\,dy \cr\bth_p \equiv &dp - r\, dx
-f(u)\, dy \cr
\bth_q \equiv &dq - f(u)\, dx - t\,dy \quad. \cr}\eqno(2.1) $$
\no We recognize 3 inequivalent subideals, $\ck_i$, each being
effective and differentially closed, with generators as follows:
$$\matrix { \ck_1  &\quad&  \ck_2  &\quad& \ck_3 \cr \cr\bth_u
\wedge dy  &\quad&\bth_u \wedge dx   &\quad&\bth_u \wedge dx \cr \cr\bth_p
\wedge dx &\quad&\bth_u \wedge dy    &\quad&\bth_u \wedge dy \cr \cr \omit
&\quad&\bth_p \wedge dx -\bth_q \wedge dy    &\quad&
\bth_p\wedge
dx \cr \cr \omit
&\omit &\quad&\quad&\bth_q \wedge dy. \cr } \eqno(2.2) $$
\no In Ref. 2, for example, the authors also distinguish 3
effective subideals, equivalent to our $\ck_1$ and $\ck_2$.
Their third subideal, as they indicate, is simply diffeomorphic to
$\ck_1$, so that they limit their discussion to these two.
They omit our ideal $\ck_3$ since  it generates
{\bf all} of ${{\cal I}}$ restricted to $Y$, except for the 1-forms.
 It might also be characterized as generating all 2-forms
annihilated on $Y$; one
might then be justified in calling it ``minimally complete''
relative to the EW procedure.\ft{38} \v
{}From the point of view of the ``classical'' symmetries,
over $J^2(M,N)$, or the higher symmetries over $J^\infty$, each
ideal has exactly the same content, as determined
 by the standard methods.\ft{39}
Nonetheless, they are properly described as
inequivalent since they engender distinct E-W coverings, as already
noted by Pirani {\it et al.\/}\ft{40},
although they did not pursue the consequences of that
inequivalence very far.
\v
Application of the E-W procedure to any of our $\ck_i$ gives the
E-W commutator equation, (1.6), in the form
$$ [{\b F} \,,\,{\bf G}] = -p\,{\bf G}_u+q\,{\b F}_u + f(u)
({\b F}_p-{\bf G}_q)\quad,\eqno(2.3a)$$
$$\hbox{with }\ {\b F} = {\b F} (u,p;w^A)\quad\hbox{ and }\quad
{\bf G} = {\bf G}(u,q;w^A)\quad,\eqno(2.3b)$$
\no  where, as is usual,\ft{41} we have ignored any possible dependence on the
independent variables,
$\{x,y\}$.  Notice explicitly that \E{2.3b} indicates that ${\b F}_q = 0
= {\bf G}_p$.  In addition to these requirements on ${\b F}$
and ${\bf G}$, the smaller ideals described in \E{2.2} cause additional
constraints on the functional dependencies
of ${\b F}$ and ${\bf G}$:\par
$$\matrix{ \ck_1&\qquad& \ck_2 &\qquad& \ck_3\cr\cr
{\b F}_u = 0 = {\bf G}_q   &\qquad&{\b F}_p + {\bf G}_q = 0
&\qquad&{\rm no \ additional \
constraints} . \cr }\eqno(2.3c) $$\par
\vskip12truept
As expected, the ideal $\ck_3$ generates a set of covering equations
identical to those of Krasil'shchik:\ft{32}  An expansion of ${\b F}$
in a Taylor series in the variable $p$, and ${\bf G}$ in the
variable $q$, leads to his equations (1.25).  These
equations are so general that \underbar{we} have been unable to find the
general solution, except when the covering space is restricted to be
only one-dimensional.\ft{42}
\v
On the other hand, the additional constraints
required by the use of either of the two smaller subideals causes
the set of equations
to become manageable. The equations we obtain, when solving the covering
equations for the set of generators, $\ck_2$, were first obtained by
Shadwick.\ft{21}
He gave the general solution for the case of 1-dimensional coverings, and then
wrote down the general \underbar{form} of the solution for multi-dimensional
coverings.
He referred to the algebra so generated as $\alpha_f$, and was apparently
only interested in cases where it had a non-trivial, homomorphic mapping to
a {\bf finite-dimensional algebra}, $\alpha$.  He considered 4 fairly simple
cases
for $\alpha$, showing that the set of functions $f(u)$ for which this gave
B\"acklund transformations depended on one's choice of $\alpha$.  We will
generalize this, determining the form of the general solution to those
equations, for arbitrary dimension.
\v
The simplest ideal, that generated by $\ck_1$, has
occurred many times in the literature, as might be expected.
Pirani et. al.\ft{13}
distinguish it by noting that it possesses a characteristic vector, $\p_q$.
As discussed in some detail in Ref. 2, those authors add on the
additional constraints that the solution for ${\b F}$ should be linear in $p$,
and indicate that ``a solution'' is obtained where the Lie algebra describing
${\b F}$ and ${\bf G}$ is simply ${\bf sl}(2,{\complx})$.
On the other hand Hoenselaers\ft{20} also requires that the solution for
${\b F}$ should be linear in $p$, but finds a rather more general
solution involving
the infinite-dimensional Lie algebra ${\bf sl}(2,{\complx})\otimes
\complx[\lambda^{-1},\lambda]$, the loop algebra over
 ${\bf sl}(2,{\complx})$.  Along the way to determining this solution,
Hoenselaers requires, additionally, of the equations that ${\bf G}_{uu}+{\bf G}
= 0$,
and that a particular central element should be ignored.  As we will in
fact obtain the most general solution to the equations, we will indicate
how these special cases are contained in them, in Section III.  The
requirement of linearity was avoided by Dodd and Gibbon, in the Appendix of
Ref. 12.  They looked for a solution for ${\b F}$ in the form of a finite
polynomial in $p$, and indicated that their structure yielded an
infinite-dimensional Lie algebra.  However, for reasons unclear to us they
only slightly generalized Hoenselaers requirement, to the requirement
${\bf G}_{uu} + {\bf G} =\b{Z}$.  Our general solution, with
infinitely many terms in the description of both ${\b F}$ and ${\bf G}$,
will also include their case.
\v  Since the requirements of $\ck_1$ are stronger than those
of $\ck_2$, surely one is embedded within the other.  The explicit
embedding is not  immediate
since $\ck_1$ is quite asymmetric, while $\ck_2$ is symmetric, from the
point of view of the first derivatives.  However, after having obtained both
general solutions, for $\ck_2$ and $\ck_1$, we will describe the
complete set of defining relations for those prolongation structures,
and show how there is a gauge transformation that transforms
a large subset of the
$\ck_2$ structure into a large subset of the one for $\ck_1$.
We will also show why both universal algebras must be infinite-dimensional,
even though we cannot explicitly identify them in terms of already-known
algebras.  From Hoenselaers' work, \ft{20,24}
we do know that the universal algebras of these
three incomplete Lie algebras (of vector fields over
the covering fibers) form a sequence:  ${\bf sl}(2,{\complx})\otimes
\complx[\lambda^{-1},\lambda]\,\equiv\,A_{1}\otimes\complx[\lambda^{-1},
\lambda]
\subseteq\qa_1 \subseteq\qa_2 \subseteq\qa_3$.
\v
One last, quite interesting ideal used for the
sine-Gordon equation should be
mentioned, even though it is in fact not a subideal of the restricted
contact module.  This is the CC ideal used by Estabrook\ft{43}, by
Harrison\ft{14},
and also espoused by Hoenselaers\ft{24}.  Such ideals were
chosen by them because of the ease with which they can be manipulated, if they
can be found.
A CC ideal is composed of two distinct sets of 2-forms,
made from some original set of 1-forms $\{\xi^a\}$ using only {\bf constant
coefficients}, thereby creating the name CC ideal.
The first set expresses the closure conditions for the $\{d\,\xi^a\}$ in
terms of wedge products of the $\{\xi^a\}$ themselves, while the second set
is simply sums of wedge products of the $\{\xi^a\}$, which may be thought of
as expressing various linear dependence conditions between them.  (It is of
course this last property that causes such ideals not to be subideals of the
restricted contact module, since all the generators there are linearly
independent.)  Estabrook came upon the study of such ideals by realizing
that they may be created by beginning with a set of Maurer-
Cartan relations for some group and annihilating some of the members ``by
hand.''  On the other hand, if one begins with a specific pde, such as our
sine-Gordon equation, one begins with some particular coordinate
representations
of the 1-forms, choosing them  so that the closure conditions may be written
with constant coefficients.  Once such an ideal is created, these authors then
``forget'' the original coordinate formulation of the 1-forms $\xi^a$, and
treat the set of closure conditions as a defining ideal for their
problem.  Typically such an ideal need no longer be a subideal of the
contact module restricted back to $Y$, nor any prolongation of it.  For the
sine-Gordon equation they choose four 1-forms, and set up their defining
relations along the lines followed earlier by Chern and Terng\ft{9}, who
interpreted
the original geometric visualizations, involving two distinct surfaces
of constant
negative curvature embedded in $\real^3$, in terms of differential geometry.
\v From our point of view
Estabrook's ideal appears rather like {\it two distinct copies} of our subideal
$\ck_1$, multiplied
with coefficients $\sin u$ and $\cos u$, respectively.  In addition, since they
use the ideal
itself to determine suitable coordinates on the manifold, this actually does
generate two copies of the sine-Gordon equation, in distinct dependent
variables;
i.e., this ideal comes pre-disposed toward an auto-B\"acklund transformation.
Taking the notion that $F^A$ and $G^A$, from \E{1.4}, should be replaced by
some set of $F^A{}_a(w)$ such that $\omega^A = -dw^A + F^A{}_a\xi^a$, one
may again find equations to determine the Lie algebraic structure generated
by these $F^A{}_a$, also usually infinite-dimensional.  In Ref. 43,
Estabrook writes explicitly a realization of the generators and known structure
constants
for this algebra; however, no one has yet been able to identify that entire
Lie algebra or place it in a universal context.  (Of course it does have a
homomorphism to ${\bf sl}(2,\complx)$, which generates the
usual B\"acklund transform.)
\v\v
\no{\bf III.~~The Algebra $\qa_1$ for the sine-Gordon Equation:}
\v Specializing directly to the explicit sine-Gordon equation,
we first consider our smallest ideal, described by the $\ck_1$ part of
\E{2.3c}, which reduces the commutator equation to the form:
$$\qa_1:\cases{\quad[{\b F}\,,\,{\bf G}] = -p\,{\bf G}_u + {\b F}_p\sin u&,\cr
\noalign{\vskip3pt}\quad{\b F}_u = 0 = {\b F}_q\,,\,
{\bf G}_p = 0 = {\bf G}_q&,\cr}\quad.\eqno(3.1)$$
\no Taking the second derivative with respect to $u$,
of the first of \Es{3.1}, we acquire
$[{\b F}\,,\,{\bf G}_{uu}] = -p\,{\bf G}_{uuu} - {\b F}_p\,\sin u$.  Summing
this one with the
original eliminates the term containing $\sin u$, so that we may first
solve $$[{\b F}\,,\,{\bf Z}] = - p\,{\bf Z}_u\qcomma\hbox{ with
}\quad{\bf Z}\equiv{\bf G}+{\bf G}_{uu}\quad.\eqno(3.2)$$
As usual, assuming that ${\b F}$ is analytic about the origin, we write
$${\b F}(p) = \sum_{m=0}^\infty {p^n\over n!}{\b F}_n
\Longrightarrow\cases{[{\b F}_n\,,\,{\bf Z}] = 0, &$n\ne1\,$,\cr
[{\b F}_1\,,\,{\bf Z}] = -{\bf Z}_u,&$n=1\,$.\cr}\eqno(3.3)$$
\no Since ${\b F}_u = 0$, this vector-field valued pde simply describes
 the `flow' of one vector
field along the direction described by another. The adjoint presentation
of such a flow is well-known\ft{44,45}; phrased in notation suitable for our
current purposes, we described it in more detail in Ref. 23.  From that
we infer
the existence of a vector field, ${\bf Q}\in W_*$, and some constraints, such
that
$${\bf Z} =\ee^{-u\>\ad{{\b F}_1}}{\bf Q}\quad,\qquad[{\b F}_n\,,\,(\ad{\b
F}_1)^m{\bf Q}]
= 0,\qcomma n\ne 1\,,\, m = 0,1,2,\ldots\quad,\eqno(3.4)$$
\no where the $ad$-operator is the usual mapping $ \ad:\,{{\cal A}}\rightarrow
{{\cal A}}$ so that $\{\ad{X}\}(Y) \equiv [X,Y]$.
\v
Having this form for ${\bf Z}$, we may solve the differential
equation for ${\bf G}$.  The homogeneous portion is
straight-forward; however, the inhomogeneous term that ${\bf Z}$
represents requires writing ${\bf G}$ also as a series
in $u$, comparing coefficients of $u^n$ on both
sides of the inhomogeneous
equation, and then re-summing the series.  The comparison gives the
recursion relation  ${\bf G}_{n+2} = -{\bf G}_n +(-\ad{\b F}_1)^n{\bf Q}$,
which has as solution the form
$$\eqalign{{\bf G} = & {\bf G}_0\>\cos u - {\bf G}_1\>\sin u +
\sum_{n=0}^\infty\,(\cos u)^{(-n-2)}(-\ad{\b F}_1)^n
{\bf Q}\quad,\cr
 = & \ {\bf G}_0 \cos u - {\bf G}_1 \sin u + \int_0^u
dw \sin (u-w) e^{-w\>\ad{{\b F}_1}}{\bf Q}\quad.\cr}\eqno(3.5a)$$
\no where the {\it negative superscripts in parentheses}, on the cosine
function, are meant
to indicate integrals, from 0 to an upper limit of $u$:
$$ \displaylines{(\cos u)^{(-p-2)} \equiv \int_0^u
du_1\>\cdots\>\int_0^{u_{i_{p+1}}}
\!du_{i_{p+2}}\,\cos u_{i_{p+2}}= \sum_0^\infty(-1)^n{u^{2n+p+2}\over
(2n+p+2)!}\cr
\hskip0.75truein =\ \cdots\ =\int_0^{u}du^\prime\,{(u-u^\prime)^p\over
(p)!}\>\sin u^\prime\cr}$$
\no Integral forms like this are also obtained by Dodd and Gibbon,\ft{22}
although they appear to have ignored the contributions from a possible lower
limit.
\v Inserting this form back into the original pde, we acquire the additional
constraints
$$[{\b F}_{pp},\,{\bf G}_0]
= 0\qcomma [{\b F}_{pp},\,{\bf G}_1] = -{\b F}_{ppp}\qcomma[{\b
F}_{pp},\,e^{-w\>\ad{F_1}}{\bf Q}]
= 0\quad,\eqno(3.6)$$
from which we obtain an explicit equation for
${\b F}$, giving us the
general solution in terms of some six generating
vector fields, $\{ {\b F}_0,
{\b F}_1, {\bf K}, {\bf G}_0, {\bf G}_1, {\bf Q}\}\in W_*$,  which elaborate
the
forms of ${\b F}$ as follows:
$$\eqalignno {  {\b F} &=
{\b F}_0 + p {\b F}_1 +  \sum_{n=0}^\infty {p^{n+2} \over
(n+2)!}(\ad{{\bf G}_1})^n {\bf K}   \cr
& = {\b F}_0 + p{\b F}_1 +
\int_0^p ds\, (p-s)\, e^{s\>\ad{{\bf G}_1}}  {\bf K}, &(3.7)  \cr }$$
\no in addition to the set of requirements that the pde makes on
the commutators of these fields.
Defining new quantities
$${\bf K}_n \equiv (\ad{G_1})^n\;{\bf K}\quad,\qquad\quad{\bf Q}_m \equiv
(-\ad{F_1})^m\;{\bf Q}\quad,\eqno(3.8)$$\no
we may write the requirements of the pde, on the
 commutators, in the following
display, where the entry inside the table is the commutator of the element
labelling the row with the element labelling the column, and the space in
the matrix is left blank if
the pde makes {\it no} requirement on that particular commutator:
$$\bordermatrix{&{\b F}_0&{\b F}_1&\b K_n&\b g_0&\b g_1&\b q_n\cr
\b f_0&0&&&0&-\b f_1&0\cr
\b f_1&&0&&\b g_1&-\b g_0 +\b q-\b k&-\b q_{n+1}\cr
\b k_m&&&&0&-\b k_{n+1}&0\cr
\b g_0&0&-\b g_1&0&0&\cr
\b g_1&\b f_1&\b g_0 +\b k-\b q&\b k_{n+1}&&0\cr
\b q_m&0&\b q_{n+1}&0\cr}\quad.\eqno(3.9)$$
\v One obtains Dodd and Gibbon's form by insisting that
our ${\bf Z}$ should be
the same as theirs, which requires insisting that our commutator,
${\bf Q}_1 = [{\bf Q}_0,{\b F}_1]$, and all higher such commutators, should
vanish.  On
the other hand, their approach allows for the existence of
our infinite sequence of vector fields, ${\bf K}_n$.
Hoenselaers' approach makes the requirement that ${\b F}$ should be linear in
$p$, so that one must set to zero
${\bf K}_0$, and all commutators of it with ${\b F}_1$.
Additionally Hoenselaers' requirements on ${\bf G}$, i.e.,
that ${\bf G}_{uu} + {\bf G} = 0$, require that we set to zero our ${\bf Q}_0$
and all
repeated commutators with ${\b F}_1$.
In our notation, Hoenselaers' mapping, $\Psi\colon \qa_1\longrightarrow
A_1\otimes\complx[\lambda^{-1},\lambda]\subset A_1^{(1)}$ is given by
$$\Psi\,\colon\ \left\{\eqalign{\b f_0 \rightarrow J^{(1)}_1\ \>, & \>\b f_1
\rightarrow J^{(0)}_3\>,\cr
\b g_0 \rightarrow J^{(-1)}_1\>, & \>\b g_1
\rightarrow J^{(-1)}_2\quad,\cr}\right.\eqno(3.10)$$
\no where the three $J_i$'s at level zero, i.e., \setify{J_i^{(0)}\equiv
J_i\,\mid\
i = 1, 2, 3}, form a basis for ${\bf sl}(2,\complx)$, with commutators
$[J_1,J_2] = -J_3\>,\> [J_2,J_3] = J_1\>,\> [J_3,J_1] = J_2$.
In order for these requirements to be consistent, we must
impose two extra restrictions on our algebra:
$0 = N = [\b f_1,[\b f_1,[\b f_1,\b f_0]]] + [\b f_1,\b f_0]$, and $
[\b g_0,[\b g_1,[\b g_1,\b g_0]]] = 0$.
\no The first of these is the statement that the central quantity, $N$,
is to be ignored.  This mapping is also pointed out by Omote,\ft{19} and is
equivalent to that specified by the use of powers of $\lambda$ in \S 6.2 of
Leznov
and Saveliev.\ft{34,46} \v\v\goodbreak
\no {\bf IV.  The algebra $\qa_2$ }\v
 For $\ck_2$ generated by the elements in
\E{2.3}, we easily see that ${\b F}$ and ${\bf G}$ must
have the following forms where ${\bf B}$ and ${\bf C}$
are as-yet-undetermined functions of $u$, which must, however, satisfy the
equations indicated:
$$\qa_2 : \left\{\matrix{{\b F} = \h p{\bf R} +
{\bf B}\quad,& \quad{\bf G} = - \h q {\bf R} + {\bf C},& \quad {\bf R}_{u} =
0\quad,
 &\cr\cr \left[ \h {\bf R}\,,\, {\bf B}\right]  = {\bf B}_{u}\quad,&
\quad \left[ \h {\bf R}\,,\, {\bf C}\right] = - {\bf C}_{u}\quad,& \quad[{\bf
B}\,,\,
{\bf C}] = {\bf R}\,f(u), &\cr }\right.\quad.\eqno(4.1) $$
\no We will usually simply refer to the arbitrary function $f(u)$ for our
equation;
however, we will shift to the special case $f(u) = \sin u$ when it becomes
necessary.
 Following other cases, such as, for
instance, the Liouville equation, or the Tzitzeica-Dodd-Bullough
equation\ft{47} would also be possible from
our general equations, but
 we have not carried it through in any detail.
\v \Es{4.1} are the same as those found by Shadwick in Ref. 15, under the
notational transformation (from our notation to his), ${\bf B} \rightarrow
{\cal A}$, ${\bf R} \rightarrow 2{\cal B}$, and ${\bf C} \rightarrow {\cal C}$.
The vector-field valued pde's in \Es{4.1} are the same sort as discussed
in the previous section.  Therefore, integration of the two differential
equations for
${\bf B}$ and ${\bf C}$ puts into evidence two new, vertical vector fields,
 ${\bf E} \,,\,{\bf J}\in W_*$, such that
$${\bf B} = e^{+{{\scriptscriptstyle{1 \over 2 }}} u(\ad {\bf R})} {\bf E},
\quad {\bf C} =
e^{-{{\scriptscriptstyle{1 \over 2 }}} u(\ad {\bf R})} {\bf J}\quad.\eqno(4.2)
$$
\no Insertion of these forms into the last of \Es{4.1} gives
the following commutator equation, which we will solve by treating it as an
identity in formal series in the variable $u$:
$$ [e^{+{{\scriptscriptstyle{1 \over 2 }}} u(\ad {\bf R})} {\bf E}\,,\,
e^{-{{\scriptscriptstyle{1 \over 2 }}} u(\ad {\bf R})} {\bf J}] = {\bf R}\>f(u)
\quad.\eqno(4.3)$$
\no Comparing coefficients of $u$ gives a countable list of
commutation relations to be satisfied,  namely
$$\eqalignno{\sum_{m=0}^k(\h)^k{k\choose m}[{\bf E}_{k-m},\,{\bf J}_m] & =
c_k\,{\bf R}\quad,
\forall \, k = 0,1,2,\ldots\quad,&(4.4a)\cr
\noalign{\hbox{where}}{\bf E}_m \equiv
(+\ad{\bf R})^m{\bf E} \qcomma{\bf J}_n&\equiv (-\ad {\bf R})^n
{\bf J}\qcomma\forall\,m,n = 0, 1, 2, \ldots\quad,&(4.4b)\cr}$$
\no and the $c_n$ are simply the constants given by
the coefficients in the power series for $f(u)$:
$$\hbox{define}\ f(u) \equiv \sum_{n=0}^\infty
{c_n\over n!}u^n\;,\qquad\quad\hbox{or}\quad
\ c_n \equiv (f(u))^{(n)}\Big|_{u = 0}\quad. \eqno(4.5)$$
\v The sum in \Es{4.4} simplifies enormously.  To see this,
 we first use the definitions in \Es{4.4},
the equation for $k=0$, and the Jacobi identity to tell us that
$$\bigl[{\bf J},\,{\bf E}_1\bigr] = \bigl[{\bf J},\bigl[{\bf R},{\bf
E}\bigr]\bigr] = \bigl[{\bf R},\bigl[{\bf J},\,{\bf E}\bigr]\bigr] + \bigl[{\bf
E},\bigl[{\bf R},{\bf J}\bigr]\bigr]
= -c_0 \bigl[{\bf R}, {\bf R}\bigr] -\bigl[{\bf E},{\bf J}_1\bigr] = \bigl[{\bf
J}_1,\,{\bf E}\bigr]\quad.$$
\no At the next level, the same sorts of manipulations yield
$\bigl[{\bf J},\,{\bf E}_2\bigr]  = \cdots =
 +\bigl[{\bf J}_1,\,{\bf E}_1\bigr]
= \cdots =
\bigl[{\bf J}_2,{\bf E}\bigr]\quad.$
\no An induction hypothesis then shows that
the value of $\bigl[{\bf E}_{k-m},\,{\bf J}_m\bigr]$ is independent of $m$,
when
 $0\le m\le k$:
$$\bigl[{\bf E}_{k-m},\,{\bf J}_m\bigr] = c_k\,{\bf R}\;,\quad\forall\>m \ni
0\le m\le k \quad.\eqno(4.6) $$
\v The original free algebra, generated by \setify{{\bf J},{\bf R},{\bf E}},
when divided by the countably infinite set of
relations in \E{4.6}, is still enormous.  We have
 so far been unable to resolve it into an
already-studied algebra.  Since, algebraically, there are really only
three generators---so that it can be said to be finitely generated---it
might well be that there are interesting realizations
to be found, for instance, among the Kac-Moody algebras.  If, however, we
desire to maintain distinct the realizations of all the \setify{{\bf E}_m,
{\bf J}_n} satisfying \E{4.6}, which does not immediately appear to be
consistent with the usual gradings of such algebras, it may in fact
require some other approach to infinite-dimensional algebras.
Surely it is true that the restriction to
only one pseudopotential reduces the problem to
the well-known homomorphism, with a
parameter, $\lambda$, onto its smallest interesting subalgebra,
${\bf sl}(2,{{\cal R}})$, taken with Chevalley basis,
$\{h,e,f\}$ such that $[h,e] = 2e$,
$[h,f] = -2f$, and $[e,f] = h$:
$$\eqalign{{\bf R}\,& \rightarrow \h(f-e)\cr
{\bf J}_n& \rightarrow\lambda^{-2}{\bf E}_n\cr}\qcomma{\bf E}_n
\rightarrow \h\lambda\cases{(-1)^{\scriptstyle{n\over 2}}h,& $n$ even$\,$,\cr
(-1)^{\scriptstyle{n-1\over 2}}(e+f),& $n$ odd$\,$. \cr}\eqno(4.7)$$
\v\v\no{\bf V.~~Gauge Transformations of the Algebras}\v
Because there appear to be several different avenues to follow at
this point, we first turn to the questions of gauge equivalence of
these algebras, seen as generated by connections over the covering with fibers
$W$.  The (usual) language
describing \E{1.1} as a {\it zero-curvature equation} has behind it
the notion that the total derivative over $J^\infty$, namely $D_a$,
when prolonged into
the covering space $J^\infty\otimes W$, should be treated as a covariant
derivative,\ft{34,48} with the additional quantities $\b X_a$ generating the
associated connection, since they prevent the
commutator of the prolonged total derivatives from vanishing, at least
until it is restricted to the variety that describes the (infinite)
prolongation of the pde being studied.  The role the
various $\b X_a$ play in these investigations requires that they be
elements of the Lie algebra of all vector fields
over $J^\infty\otimes W$, preserving the fibered structure.  This
makes it clear
that $\Gamma \equiv \b X_a\>dx^a$ is in fact a
 Lie-algebra-valued connection 1-form for this covariant derivative,
$\nabla \equiv dx^a\{D_a + \b X_a\}$; therefore our $\Gamma$ should
transform in the usual way for connections.
\v
The transformations we consider correspond to flows of the
covering space generated by particular tangent vector fields, so that
we are simply moving along a congruence of curves from one point on the
manifold to another.  By restricting
our attention to vertical vector fields, i.e., those
of the form of the $\b X_a = X_a^A\p_{w^A}$, any individual curve, within
the congruence, simply runs within a single fiber over its base point, in
$J^\infty$, at which it began.
Therefore, different positions along such a congruence
of curves, i.e., different values of a parameter along the curves, just
correspond to different values of fiber coordinates \setify{w^A}
over the same base point.  We surely want the structure of
our theory to be independent of distinctions such as this;
 therefore, we refer to
such transformations as {\bf gauge} transformations, i.e., transformations
which leave invariant the meaning of the equations presented, only
mapping different explicit presentations of
the underlying geometry into one another.
\v
 We now describe explicitly
 such a gauge transformation, generated by ${\bf R}$, say,
a vertical vector field defined over some (local)
portion of our manifold. The flow of that vector field
is a (local) mapping of the manifold into itself, that can be presented
via a congruence of curves, described by
$\Phi_t\equiv e^{t{\bf R}}:U\subseteq M\rightarrow M$.
As the base point is the same for all points on this curve, the actual
coordinates along the curve can be specified by first giving the
coordinates, $z_0 \in J^\infty$,
 for that base point and then integrating the
flow equations for the $w^A(z_0,t)$:
$${dw^A\over dt} = {d\over dt}\Phi_t= {d\over
dt}\left\{tR^A(z_0,w^B)\right\}\Phi_0\quad.\eqno(5.1)$$
The induced mapping of the tangent bundle,
$(\Phi_t)_*:U_*\rightarrow M_*$,
may be presented in the form ${\bf X}_P\longrightarrow
e^{t(\ad{\bf R})}{\bf X}_{\Phi_{{}_t}{}_{\scriptscriptstyle(P)}}$, for $P\in
U\subseteq M$,
and ${\bf X}_P$ an arbitrary tangent vector at that point.
Under this flow a covariant derivative operator,
$\nabla \in M_*\otimes M^*$,
with a Lie-algebra-valued connection 1-form $\Gamma$,
would have a transformation law usual for connections:
$$\Gamma_t \equiv e^{t(\ad{\bf R})}\Gamma - d(t\,{\bf R})\quad.\eqno(5.2)$$
\no In our case, the derivation fills in the steps in the following
reasoning:
$$\eqalign{(D_a^\prime +\b X_a^\prime)_\pse
& =(\nabla^\prime_a)_\pse
 \equiv e^{t(\ad{\bf R})}\{{\nabla_a}_P\}
= e^{t(\ad{\bf R})}\{(D_a + \b X_a)_P\}\cr  = &\{D_a -
({\bf R})_a\}_{e^{t{\bf R}}{\scriptscriptstyle P}} + \{e^{t(\ad{\bf R})}\b
X_a\}_\pse
= \{D_a - (t{\bf R})_a + e^{t(\ad{\bf R})}\b X_a\}_\pse
\quad,\cr}$$
\no where the primes indicate the transformed objects, and the
subscript $a$ on $(t{\bf R})_a$ is only a shorthand for $\{D_a(t{\bf R})\}$.
As well we have used explicitly the fact that the invariance of the base
manifold, $J^\infty$, under these transformations leaves the underlying total
derivatives, $D_a$, invariant.  Suppressing the manifold points, the result may
then be re-stated as simply
$$ (\Phi_t)_*(\b X_a)\ \equiv\ \b X_a^\prime = e^{t(\ad{\bf R})}\b X_a\ -
\{D_a(t{\bf R})\}
\quad,\eqno(5.2^\prime)$$
\no with the same content as \E{5.2}.
\v
Applying these notions to our current considerations, we must modify our
notation slightly, so that we can discuss both prolongations at once.  We do
this by first changing from the symbols \setify{\b F, \b G} to new ones
\setify{\cal F,\,\cal G},
for this section only, so that we may then further label separately the
connection
forms \setify{{\cal F}_2, {\cal G}_2}, as the ones determined by $\qa_2$, via
\Es{4.1-2},
and the other connection
forms \setify{{\cal F}_1, {\cal G}_1}, determined by $\qa_1$ via \Es{3.5,7}:
$$\openup-1\jot\eqalign{
\qa_2\;:\quad&\ \left\{\eqalign{{\cal F}_2 & = \h\,p\,{\bf R} +
e^{+{{\scriptscriptstyle{1 \over 2 }}} u(\ad {\bf R})} {\bf E}\,,\cr
{\cal G}_2 & =
-\h\,q\,{\bf R} + e^{-{{\scriptscriptstyle{1 \over 2 }}} u(\ad {\bf R})} {\bf
J}\,,\cr}\right.\cr\cr
\qa_1\;:\quad&\ \left\{\eqalign{{\cal F}_1 & = {\bf F}_0 + p {\bf F}_1 +
\sum_{n=0}^\infty {p^{n+2} \over
(n+2)!}(\ad{{\bf G}_1})^n {\bf K}\,, \cr
 {\cal G}_1 & =
{\bf G}_0\>\cos u - {\bf G}_1\>\sin u + \sum_{n=0}^\infty\,(\cos
u)^{(-n-2)}(-\ad{\bf F}_1)^n{\bf Q}\,.\cr}\right.\cr}\eqno(5.3)$$
\v The form of ${\cal F}_2$ immediately suggests a gauge transformation with
the
group element $e^{-{{\scriptscriptstyle{1 \over 2 }}}
 u{\bf R}}$.  The result, via \E{5.2}, is
$$ (\Phi_{-{\scriptscriptstyle{1\over 2}}u})_*\left\{{\cal F}_2,
{\cal G}_2\right\} = \left\{
{\cal F}_2^\prime  = p\,{\bf R} + {\bf E}\,,\;
{\cal G}_2^\prime  = e^{- u(\ad {\bf R})} {\bf J}\right\}\,,\eqno(5.4)$$
\no which resembles the Hoenselaers' form of the prolongation achieved via
our algebra $\qa_1$, as discussed in Section III.  Comparing this equation for
${\cal F}_2^\prime$ with our
general equation for ${\cal F}_1$, we see that the equality of ${\cal F}_2$
and ${\cal F}_1$ requires
$\b K_0 = 0$, along with its repeated commutators.
  Although one must still also check that the required
commutation relations are the same, it turns out that they indeed are.
On the other hand, the two forms for ${\cal G}$ appear different; however, by
inserting the quantities in
\E{5.4} into \Es{3.1}, we find that they do constitute a solution of
the general equations obtained from $\qa_1$.    Since ${\cal F}_2^\prime$ and
${\cal G}_2^\prime$ do
satisfy the original $\ck_1$ equations, they must also constitute
a special case of the general solution, \setify{{\cal F}_1, {\cal G}_1} as
given
in \E{5.1}  The truth of this may be seen by
comparing the fairly-simply-displayed infinite series
 for ${\cal G}_2^\prime$ in \E{5.4} with
the infinite sum for ${\cal G}_1$ in \E{3.5}.  Since all the repeated cosine
integrals, $(\cos u)^{(-n-2)}$, discussed near \E{3.5}, may be
written as either $\sin u$ or $\cos u$, plus finite polynomials in $u$,
we may expand the trigonometric functions in infinite series and
re-collect the expression for ${\cal G}_1$ in \E{5.3} in the form
$${\cal G}_1 = {\bf G}_0 - u({\bf G}_1) + {u^2\over 2!}(\b Q_0 - {\bf G}_0) -
{u^3\over 3!}
\left(Q_1 - G_1\right) + {u^4\over 4!}\left(Q_2 - Q_0 + G_0\right)
-{u^5\over 5!}\left(Q_3 - Q_1 + G_1\right) + \cdots$$
Since ${\cal G}_2^\prime$, in \E{5.4}, is already of the form of an infinite
power series, we may identify completely the general forms of ${\cal
G}_2^\prime$
and ${\cal G}_1$, allowing us to write a full presentation for an
identification mapping of one to the other:
$$\Xi\;:\;(\Phi_{-{\scriptscriptstyle{1\over 2}}u})_*(\qa_2)
\rightarrow(\qa_1)_{\big|_{{\bf K}=0}}\!
:\ \left\{\eqalign{{\bf R}\rightarrow {\bf F}_1\,,&\quad\b E_0\rightarrow{\bf
F}_0\,,\cr
\quad\b J_0\rightarrow {\bf G}_0\,,\quad\b J_1\rightarrow  & -{\bf G}_1\,,\quad
\b J_2\rightarrow \;{\bf Q}_0 - {\bf G}_0\,,\cr
\quad\b J_3\rightarrow {\bf Q}_1 + {\bf G}_1\,,\;\;&
\ldots\;,\;\b J_j\rightarrow \;{\bf Q}_{j-2} - \b
J_{j-2}\quad.\cr}\right.\eqno(5.5)$$
After also checking that {\bf all} the required commutators, from \Es{3.9}
and \E{4.4}, do indeed map into each other, these comparisons tell us that
the general algebraic solution for the prolongation structure defined by
$\ck_2$, i.e., $\qa_2$, is actually gauge equivalent to a subalgebra of
the one defined by $\ck_1$, which we believe to be unexpected.  (The subalgebra
in question is obtained from $\qa_1$ by setting the
generator $\b K_0$, and therefore all the other $\b K_n$'s, to zero.)

\v An alternative approach to
gauge transformations of this sort is to view them as simply effecting
a different set of choices for the fiber coordinates $w^A$.  This is a
reasonable viewpoint since the vector fields generating the flows are always
vertical.   Unlike the
previous calculations, this approach is not as abstract; however, it
will require an explicit realization of the vector
field ${\bf R}$, in order to integrate \Es{5.1}.  Equations (1.4)
tells us that an acceptable solution to the problem allows us to
interpret the components of ${\cal F}_2$ and ${\cal G}_2$ as first derivatives
of
the pseudopotentials. In order to distinguish the two, in-principle-distinct
sets of pseudopotentials, we use \setify{w^A} for those from $\qa_2$ and
\setify{v^A} for those from $\qa_1$:
$${\cal F}_2^A = (w^A)_x\,,\;{\cal G}_2^A = (w^A)_y\,,\qquad
{\cal F}_1^A = (v^A)_x\,,\;{\cal G}_1^A = (v^A)_y\,.\eqno(5.6)$$
\no Beginning with any
particular solution to the prolongation problem for $\ck_2$, a plausible (and
usual) realization for ${\bf R}$ is simply $2\p_{w^1}$, as done, for instance,
by
Shadwick.\ft{21}   This choice realizes our gauge transformation as the flow
presented
as $e^{-p\p_{w^1}}$,  making the solutions to the flow equations
very simple.  The curves of the congruence for the gauge transformation
simply go along the direction of increase of $w^1$, from the (arbitrary)
original value, by an amount $-p$, the jet variable for $-u_x$; i.e., we have
$${w^1}^\prime = w^1 - p\,,\hskip0.6truein
\hbox{all other coordinates stay the same.}\eqno(5.7)$$
\v
Using this realization for ${\bf R}$ the $\qa_2$ portion of \Es{5.3} can
be re-written as
$$\eqalign{(w^A)_x & = u_x\,\delta^A_1 + \left\{e^{u\p_1}\b E\right\}^A
   = u_x\,\delta^A_1 + \sum_{n=0}^\infty{u^n\over
n!}(\p_{w^1})^nE^A(w)\quad,\cr
(w^A)_y & = -u_y\,\delta^A_1 + \left\{e^{u\p_1}\b J\right\}^A
  = -u_y\,\delta^A_1 + \sum_{n=0}^\infty{(-u)^n\over n!} (p_w^1)^nJ^A(w)
\quad.\cr}\eqno(5.8)$$
\no
Writing everything at the transformed point, \Es{5.7} give us the
gauge-transformed
realization:
$$\eqalign{({w^A}')_x & = (1+\delta^A_1)\,u_x + E^A\quad, \cr
({w^A}')_y & = (-1 + \delta^A_1)\,u_y + \sum_{n=0}^\infty {(-2u)^n\over n!}
(\p_{w^1})^nJ^A\quad.\cr}\eqno(5.9)$$
\no
On the other hand, using the
identification of ${\cal F}_1$ with ${\bf R}$ and the realization of ${\bf R}$
already
agreed to, we re-write the $\qa_1$-equations, with $\b K = 0$~:
$$\eqalign{(v^A)_x & = 2 u_x\delta^A_1 + F_0^A\quad,\cr
(v^A)_y & = G_0^A\,\cos u - G_1^A\,\sin u + \sum_{n=0}^\infty
(\cos u)^{(-n-2)}(-2\p_{w^1})^nQ^A\quad.\cr}\eqno(5.10)$$
\v Comparing \Es{5.9} and (5.10), and recalling that we have already shown
that the two forms of ${\cal G}$ are equivalent, we see that the two equations
show that ${w^1}'$ is the same as $v^1$, although the other fiber coordinates
differ substantially in the two presentations.
This difficulty lies only with the
particular choice for a realization for ${\bf R}$, which
works well for the 1-dimensional situation, but must be generalized for a
higher-dimensional case.  For instance,
a realization of the form $R^A = 2$, for {\bf all} allowed
values of $A$, would arrange it so that {\bf each} of the
transformed variables \setify{{w^A}'} would simply be a translation of the
un-transformed ones, and would then correspond to the variables \setify{v^A},
appropriate to the other algebra.
This would {\bf realize} the gauge equivalence of ${\cal K}_2$ with that
subalgebra of ${\cal K}_1$ that has set to zero $\b K_0$ and its repeated
commutators with ${\bf F}_1$, with the meaning that the origin in
\setify{w^A}-space
must be chosen differently for different presentations of the prolongation
structure.  This result might easily be expected for objects like potentials!
\v Returning to the 1-dimensional case, it is well known in that
case\ft{21} that the general solution will not permit non-zero values for
either $\b Q$ or $\b K$.  Therefore, when restricted to 1-dimensional
fiber spaces, the two algebras are completely equivalent.  In a less
formal mode of expression, this fact has been well-known for quite some
time.  This gauge equivalence, when phrased simply as a change in
the names of the variables, so that one replaces $w$ by some $w^\prime + u$,
has long been known\ft{49} as a simple way of beginning, say, with the
simplest ideal of 2-forms, our $\ck_1$, and nonetheless recovering
exactly the 19th century version of the B\"acklund auto-transformation
that comes out immediately if, instead, one uses the ideal $\ck_2$ and
restricts oneself to 1-dimensional fibers.\ft{21}  This last transformation
only works, of course, provided the same (1-dimensional) realization of ${\bf
sl}(2,\complx)$
has been chosen for the generators of the prolongations coming from the
two ideals.
\v
Our understanding of gauge transformations as relevant to EW prolongations
allows us to also return to our work on Burgers' equation and the KdV
equation, in Ref. 23, where we determined the continuation
of those prolongation structures to higher-level jets.  We noted then that
the geometric meaning was still unresolved.  Now it is clear that the
exponential operators in \Es{4.10-14}, for Burgers' equation, may all be
removed.  Beginning with \E{4.10} there, the extra generator $\b A$ is
really simply a gauge freedom for the origin of coordinates in the fiber
space $W$.  A gauge transformation of the form $e^{z(\ad{\b A})}$ will
remove the exponential in front of the expression, and also the additional
term $z_1\b A$, so that this new choice of $w^B$'s, over $J^2$,
retains the same expression for the $X_a$ as they had over $J^1$.  This
process may be continued as one allows the EW prolongation to be defined
on higher and higher jets.  Bringing the connection from, say, $J^m$ to
$J^{m+1}$, a gauge transformation $e^{z_{m-1}\ad{\b A}}$ will retain
invariant the connection, with this
amounting only to the transformation of fiber coordinates, $w^B \ \rightarrow
w^B + z_{m-1}A^B(w) + \,\ldots\,$, so that no interesting new structure is
in fact added.  On the other hand, when the EW prolongation for the KdV
equation, in \Es{4.28} and (4.34), there, is gauge transformed in the same
way, they take on a much simpler appearance, but it is clear that they
nonetheless retain the important new structures that are identified there
as generators for the higher (local) symmetries of the KdV equation.  The
process of defining the nonlocal EW prolongation on a higher jet adds to
it explicitly new generators which carry the information concerning the
higher local symmetries that are defined on the corresponding higher jet.
\v\v\goodbreak
\no{\bf VI.~~Further Attempts to Identify $\qa_2$}\v
\v We have been quite involved in attempts to seriously identify the entirety
of
the algebra $\qa_2$.  While some partial results are presented below, the
algebra
has so far resisted its identification in terms of objects with which we are
familiar.  Therefore, we present these results in the hope that other
researchers
interested in such results may be able to utilize them to proceed further.  As
a
beginning, we may utilize the
gauge transformation from the previous section to transform our previous
homomorphism, $\Psi\>:\>\qa_1\ \rightarrow A_1
\otimes\complx[\lambda^{-1},\lambda]$, given
by \Es{3.10}, into a homomorphism for $\qa_2$, namely
$\Psi'\equiv \Psi\circ(\Phi_{-{\scriptscriptstyle{1\over 2}}u})_*
: \qa_2 \longrightarrow A_1\otimes\complx[\lambda^{-1},\lambda]$,
 which may lead to some additional insight concerning
more general mappings and identifications for that algebra. The first step in
a complete description for $\qa_2$ lies
in a determination of the commutators of the $\b E_i$ with themselves and
the commutators of the $\b J_k$ with themselves.
Beginning with $\Xi$, from \Es{5.5}, we may construct:
$$ \b E_m \equiv (\ad {\b R})^m\b E\ \mathop{\longrightarrow}_\Xi
\ (\ad {\b F}_1)^m\b F_0\ \mathop{\longrightarrow}_\Psi\
\cases{(-1)^{{m-2\over 2}}\>J_1^{(1)},& for m even,\cr
(-1)^{{m-1\over 2}}\>J_2^{(1)},& for m odd,\cr}\eqno(6.1)$$
\no and then continue by attempting to construct the double commutators $[\b
E_m,\b E_n]$.
Under the action of $\Xi$ itself, these undetermined commutators remain
undetermined.  However, $\Psi'$ does indeed cause these to be determined:
$$[\b E_m,\b E_n]\ \mathop{\longrightarrow}_{\Psi'}\ \cases{0\,,& $m + n$
even,\cr
(-1)^{{m+n-1\over 2}}\,J_3^{(2)}\,,& $m + n$ odd.\cr}\eqno(6.2)$$
\v
 It is clear that triple
commutators of the $\b E_m$'s among themselves will generate elements
of $A_1^{(1)}$ at the third level, quadruple commutators will generate elements
at the
fourth level, etc.  The situation for the original $\b J_k$'s is slightly
different
in the sense that the `extra' elements, $\b Q_j$, are also involved, as
described in
\Es{5.5}.  However, none of the commutators involving only elements from the
set \setify{\b G_0, \b G_1, \{\b Q_j\}} are determined within the general form
of $\qa_1$.
Therefore, the final mapping, $\Psi'$ that sends them into $A_1^{(1)}$ sends
them in
exactly the same way as it does the $\b E_m$'s, except that the result lies at
negative levels of the structure of $A_1^{(1)}$, rather than positive levels.
 In this way, the entire structure of $\qa_2$ fits very
nicely into the loop algebra that is $A_1^{(1)}$ without its center.
This homomorphism does, however, lose quite a lot of information that might
have been
contained within the larger algebra.  For example, we see that $\Psi'(\b
E_{m+2}) =
-\Psi'(\b E_m)\,,\Psi'(\b J_{k+2}) = -\Psi'(\b J_k)$, so that those particular
(countably-) infinite strings of generators are reduced to only 4.  As well,
the mapping loses all the information that might be in the original generators
$\b Q_i$ and $\b K_j$, since it also relates $\Psi'(\b E_m)$ and $\Psi'(\b
J_m)$
simply by the change of sign of the index $m$; when the
loop-algebra elements are realized by powers of a ``spectral,'' or loop
parameter, $\lambda$,
this simply maps $\lambda^m$ into $\lambda^{-m}$.
\v As described in Section I, we still feel that a better understanding of
these
algebras would benefit our understanding of the general solutions of the pde's
involved.  There we insisted
 that (at least) those generators involved explicitly in the description
of ${\bf F}$ and ${\bf G}$ should remain linearly independent, which
requirement
 is violated by
the mappings $\Psi$ and $\Psi'$,  that
map the infinite
sequence of, for instance, the $\b E_m$'s into just two unique ones, $\Psi'(\b
E_0)$
and $\Psi'(\b E_1)$, the rest just being $\pm 1$ multiplied by these two.
Therefore, we have searched for other
realizations that would be faithful for at least the entire set  $\{{\bf
J}_n\,,\,{\bf R}\,,\, {\bf E}_m\,|\,m,n = 0, 1, 2,\ldots\}$.
Unfortunately, we have been unable to find such a faithful realization.  It
seems
necessary to make some additional requirements on the algebra, still of course
preserving the requirement of faithfulness.  The
most reasonable alternative we know was
suggested by Alice Fialowski, namely that we should restrict ourselves to
the subalgebra which is
spanned, in the sense of a vector space, on the countable
list of generators presented just above.
This is again quite distinct from the homomorphisms, $\Psi$ and $\Psi'$, since
they
send multiple commutators of the $\b E_m$'s
with themselves into {\bf new quantities}, corresponding to larger powers of
the loop
parameter $\lambda$, which are {\bf not within the span} of the original
generators.
Nonetheless, we do believe that there is some merit in continuing this path for
awhile.  As we will show, it leads to interesting connections with Banach
algebras of
Toeplitz-type operators, a different direction than has so far been explored by
researchers studying symmetry groups for pde's.
\v
To describe the algebraic restrictions on $\qa_2$ generated by insisting
that the generators should be linearly independent and
should span the (vector) space,
we could require that the double commutators, $[\b E_m,\b E_n]$ and $[\b J_m,\b
J_n]$
be given as linear combinations of the original set of generators.
 The general form of the requirements on all these coefficients, as
required by the satisfaction of the Jacobi relations, is rather large and
complicated.
However, one property may be quickly seen.  If all the coefficients in the
directions $\b J_k$ of the
commutators $[\b E_m,\b E_n]$ are zero, {\bf or} all the coefficients in the
directions $\b E_m$ of the commutators $[\b J_j,\b J_k]$ are zero, then it is
impossible to maintain the linear independence of all the original generators.
 Therefore, once again we make yet one more `extra' assumption,
going toward some notion of linear independence which is clearly not the most
general,
but which did at least arrive at some unexpected places.
\v We therefore now discuss, in considerable detail, the situation where one
assumes that the commutators are indeed
spanned only by those coefficients
essential for linear independence.  More precisely, we consider the particular
special subalgebra of $\qa_2$, which we characterize as $\qb_2$, which contains
as
a vector space basis the generators
$$\{{\bf J}_n\,,\,{\bf R}\,,\, {\bf E}_m\,|\,m,n = 0, 1, 2,\ldots\},\ \hbox{
a vector space basis for ~}\ \qb_2\quad.\eqno(6.3)$$
\no As this is the
most general case we will consider in any detail, it seems reasonable to
recapitulate
here some of the earlier relations as well so that we may collect together the
entire set of relations that define this subalgebra:
$${\qb_2}\quad\colon\quad\left\{\eqalign{
{\bf E}_m \equiv
(+\ad{\bf R})^m{\bf E} \qcomma{\bf J}_n&\equiv (-\ad {\bf R})^n
{\bf J}\qcomma\forall\,m,n = 0, 1, 2, \ldots\quad,\cr
\bigl[{\bf E}_{k-m},\,{\bf J}_m\bigr] & = c_k\,{\bf R}\;,\quad\forall\>m \ni
0\le m\le k \quad,\cr
\bigl[{\bf E}_i,{\bf E}_j\bigr] = A_{ij}{}^m{\bf J}_m\quad
,&\quad \bigl[{\bf J}_m,{\bf J}_n\bigr] =
B_{mn}{}^i\,{\bf E}_i\quad ,\cr}\right. \eqno(6.4)$$
\no where the scalar quantities, $A_{ij}{}^m$ and $B_{mn}{}^i$, are
the needed coefficients already mentioned.
Applying the Jacobi relations to this set of definitions determines a
complete set of constraints on them which we present below.  In addition,
we find that they offer an unexpected expression as
the defining equations for an algebra of Toeplitz
operators, although we do not yet have explicit realizations.
\v To describe the constraints on these coefficients,
it is more useful to consider each of them as being the elements of a single
countable sequence of (semi-infinite by semi-infinite) matrices, that of course
may be taken as the adjoint representation of our algebra.  We
collect these coefficients into matrices, ${{\cal A}}_i$ and ${{\cal B}}_m$, by
the rules that
$$({{\cal A}}_i)^m{}_j \equiv A_{ij}{}^m \qcomma ({{\cal B}}_m)^i{}_j \equiv
B_{mj}{}^i\quad,\quad
i,j,m = 0, 1, 2, \ldots\quad,\eqno(6.5)$$
\no To manipulate the elements, it is useful to have the
usual shift, or `creation' matrix, $\Lambda$, with each element $+1$ in its
first ``super-diagonal,'' and all other elements zero, together with its
transpose, the `destruction' matrix, $\Lambda^T$:
 $$ \Lambda^j{}_i = \delta^{j+1}_i\,,\; (\Lambda^T)^j{}_i = \delta^j_{i+1}\,,\;
\Lambda\Lambda^T = I\,,\;\Lambda^T\Lambda = I - (\vec e_0)^0\,,\quad \vec
e_{i+1} =
\Lambda^T\,\vec e_i = (\Lambda^T)^{i+1}\vec e_0 \quad,\eqno(6.6)$$
\no where  $\{\vec e_i\mid i=0,1,\,\ldots\;\}$ are the usual basis vectors
which have the entry $+1$ at the i-th row and zeros elsewhere.  We also
define a sequence of vectors, $\vec {{\cal C}}_j$, with elements determined by
the constants
$c_i$, from the derivatives of $f(u)$, at $u=0$, as at \E{4.5}, which we take
to be
$\sin u$ from now on:
$$\vec C_j \equiv (\Lambda)^j\,C_0\,,\quad
 (\vec C_i)^j = (\vec C_0)^{i+j} \equiv c_{i+j}\quad.\eqno(6.7)$$
\v We may now detail the requirements which are placed on
valid matrices ${{\cal A}}_i$ and ${{\cal B}}_j$ by virtue of the
Jacobi identity.  We list them in some detail, and order, since their
solutions will require that order.  The {\bf zero-th requirement} is
simply that of skew-symmetry, which is required by their definition in terms of
commutators:
$$({{\cal A}}_i)^j{}_k + ({{\cal A}}_k)^j{}_i = 0 \qqcomma({{\cal B}}_i)^j{}_k
+ ({{\cal B}}_k)^j{}_i = 0\quad.\eqno(6.8)$$
\no Assuming the satisfaction of \Es{6.6}, the {\bf first requirement} may be
formed as a first-order recursion relation for the matrices ${{\cal A}}_i$:
$${{\cal A}}_{i+1} = -\Lambda^T {{\cal A}}_i - {{\cal A}}_i \Lambda^T \equiv
-\{\Lambda^T,{{\cal A}}_i\}\qcomma{{\cal B}}_{i+1} = -\Lambda^T {{\cal B}}_i -
{{\cal B}}_i \Lambda^T \equiv
-\{\Lambda^T,{{\cal B}}_i\}\quad ,\eqno(6.9a)$$
\no where the braces are used to indicate an {\it anti-commutator} of the
matrices. This relation can immediately be iterated to give
$${{\cal A}}_i = (-1)^i\sum_{p=0}^i{i\choose p}(\Lambda^T)^{i-p}{{\cal
A}}_0(\Lambda^T)^p\,,
\ \hbox{or}\ ({{\cal A}}_i)^k{}_\ell = (-1)^i\sum_{m=0}^i{i\choose
m}\left({{\cal A}}_0\right)^{k-i+m}
{}_{\ell+m}\quad,\eqno(6.9b)$$
\no with the same form for the ${{\cal B}}_i$, so that only the two
matrices ${{\cal A}}_0$ and ${{\cal B}}_0$ are needed.
\v
The {\bf second requirement} is an ``inverse'' eigenvector equation,
since in this case we
are given the eigenvectors and eigenvalues and must find the
associated matrices. It is also reminiscent of
group representations on vector-valued operators:
$$\vec C^T{}_{[j}\,{{\cal A}}{}_{i]} = \sum_{k=0}^\infty ({{\cal
A}}_i)^k{}_j\vec C_k^T \quad ,\quad \vec C^T{}_{[j}\,{{\cal B}}{}_{i]} =
\sum_{k=0}^\infty ({{\cal B}}_i)^k{}_j\vec C_k^T\quad,  \eqno(6.10)$$
\no where the brackets around indices indicate a ``commutator'' on the
indices, i.e.,
$X_{[i}Y_{j]} \equiv X_iY_j - X_jY_i$.
At this point the requirements on the ${{\cal A}}_i$'s and the ${{\cal B}}_m$'s
have been of
identical type. However, the
{\bf third requirement} establishes quadratic relations among them, which
should serve to distinguish the two sets:
$$\displaylines{\hfill {{\cal A}}_n{{\cal B}}_p = {{\cal B}}_n{{\cal A}}_p =
\vec e_{p+1}\,\vec C_n\!{}^T - c_{n+p}\Lambda^T =
\Lambda^T\{\cq_{np} - c_{n+p}I\}\ ;\hfill\llap{(6.11)}\cr
 \cq_{np}\equiv \vec e_p\,{\vec C_n}^T\>, \ \Longrightarrow \
\hbox{trace}\,(\cq_{np}) = c_{n+p}\quad.\cr}$$
\v These matrix
requirements, except for the one concerning skew-symmetry, may also be very
naturally
expressed via generating functions, with the recursion relation having an
especially
simple form:
$$\displaylines{\hfill\sum_{i=0}^\infty {t^i\over i!}{{\cal A}}_i =
e^{-t\Lambda^T}\!{{\cal A}}_0\>e^{-t\Lambda^T}
\quad,\hfill\llap{(6.9$^\prime$)}\cr
\hfill
\sum_{j,k=0}^\infty {s^jt^k\over j!k!}(\vec C_j^T\,{{\cal A}}_i - \vec
C_i^T\,{{\cal A}}_j)  =
C_0^T\left\{e^{-(s-t)\Lambda^T}\!\!{{\cal A}}_0e^{-s\Lambda^T} -
e^{+(s-t)\Lambda^T}
\!\!{{\cal A}}_0
e^{-t\Lambda^T}\right\}\quad,\hfill\llap{(6.10$^\prime$)}\cr
\hfill e^{-t\Lambda^T}\!\!{{\cal A}}_0 e^{-(s+t)\Lambda^T}\! {{\cal B}}_0
e^{-s\Lambda^T} =
\{e^{s\Lambda}\vec {{\cal C}}_0(\vec e_0)^T e^{t\Lambda} - \sin(s+t)I\}\Lambda
\quad.\hfill\llap{(6.11$^\prime$)}\cr}$$
\no These relations suggest that the structures are `tightly' defined
and that, perhaps, determination of the correct form for those at level 0 would
in
fact determine correctly all the others.
\v
The general solution to requirements zero and one puts
non-trivial constraints on the matrices, with the complete derivation of
these constraints being relegated to the Appendix. The general form is
determined by the fact that skew symmetry requires that $({{\cal A}}_i)^k{}_i$
must
vanish.  Iteration of this requirement provides the relationships
$$\left({{\cal A}}_0\right)^k{}_{2i}   = -\sum_{m=0}^{i-1}{i\choose
m}\left({{\cal A}}_0\right)^{k-i+m}{}_{i+m} = -\sum_{n=1}^i{i\choose
n}\left({{\cal A}}_0\right)^{
k-m}{}_{2i-n}\quad,\eqno(6.12)$$
\no constituting an $i$-term recursion relation for the elements of the $2i$-th
column.  These relations determine the even-numbered columns in
terms of the odd-numbered ones, which are still arbitrary.
To describe this result, we first conceive of the columns as vectors,
introducing two new symbols $\vec a_j$ and $\vec A_{2m}$ to describe the odd-
and
even-numbered columns of ${{\cal A}}_0$, respectively, with like objects for
${{\cal B}}_0$:
$$(\vec a_j)^k \equiv ({{\cal A}}_0)^k{}_{2j+1}\;,\ (\vec A_{2m})^k \equiv
\,({{\cal A}}_0)^k{}_{2m}\;,\eqno(6.13)$$
\no which allows the explicit description
$$\eqalign{\vec A_0 = \vec 0\qcomma\vec A_2 = -\Lambda^T\vec
a_0\quad,&\quad  \vec A_4 = -2\Lambda^T\vec a_1 + (\Lambda^T)^3\vec
a_0\quad,\cr
\vec A_6 = -3\Lambda^T\vec a_2 + 5(\Lambda^T)^3\vec a_1 - 3(\Lambda^T)^5\vec
a_0\quad,&\;\ldots\;,\;\vec A_{2i} = \sum_{n=0}^{i-1}\cp_{i, n}
\big\{(\Lambda^T)^{2n+1}\vec a_{i-n-1}\big\}\,,}\eqno(6.14)$$
\no with completely similar representations for the columns of ${{\cal B}}_0$,
and the (integer) coefficients $\cp_{i,n}$ are of alternating sign, and are
determined as finite sums of binomial coefficients, given explicitly in the
Appendix.
 \v
The second set of requirements may be explicitly solved as well.   To
demonstrate
the solution, we must
first define the alternating sums of even and of odd elements for the
odd columns, $\vec a_i$:
$$R_{j}\equiv\sum_{m=0}^\infty (-1)^{j+m}(\vec a_j)^{2m} = \vec C_1\cdot\vec
a_j \quad,\quad
S_{j}\equiv\sum_{m=0}^\infty (-1)^{j+m}(\vec a_j)^{2m+1}=
\vec C_0\cdot\vec a_j\quad,\eqno(6.15)$$
with entirely similar quantities, $U_j$ and $V_j$, respectively,
being defined for the vectors $\vec b_k$.
\no The recursion relations induce $(n+2)$-term
recursion relations for these sums, which may be iteratively solved, in terms
of
$R_0$, $R_1$, and $S_0$:
$$\eqalign{R_{m} & = w_m(R_0 + R_1) + R_1\>,\forall\,m\ge 2 \>,
 \quad w_m =
4, ~20, ~84, ~340, ~1,364, ~5,460, ~21,844,\ \ldots \>,\cr
S_{j} & = q_jS_0\>,\forall\,j\ge 1\;,
\quad q_j = 3, ~11, ~43, ~171,
{}~2,393, ~2,731, ~10,923, ~43,691,\ \ldots \>,\cr}\eqno(6.16)$$
\no with, again, identical relations for $U_m$ and $V_n$.
The calculated coefficients, $w_m$ and $q_j$,
increase very rapidly; therefore, we have made the {\it additional
hypothesis\/} that
all of $R_0+R_1$, $S_0$, $U_0+U_1$ and $V_0$ must vanish, which satisfies the
equations
without the need of allowing these divergent sequences of constants, with the
result
that all of these sums are explicitly required to have the following values:
$$\eqalign{\vec C_1\cdot\vec a_i \equiv R_i  & =
(-1)^i R_0 \equiv (-1)^i\vec C_1\cdot\vec a_0\qcomma \vec C_0\cdot\vec a_i
\equiv S_i = 0 \quad;\cr
\vec C_1\cdot\vec b_j \equiv U_j & = (-1)^j
U_0 \equiv(-1)^j \vec C_1\cdot\vec a_0\qcomma \vec C_0\cdot \vec b_j
\equiv  V_j = 0 \quad,\cr}\eqno(6.17)$$
\no quite analogous to the fact that for each value of $p$, we always have
$\sum_{j=0}^\infty (\vec C_p)^j = 0$.
\v
With the even columns of ${{\cal A}}_0$ and ${{\cal B}}_0$ determined by
\Es{6.12} and the odd columns
restrained (slightly) so that their sums satisfy \Es{6.15},
all the independent conditions have been satisfied, and now only
the quadratic equations, \Es{6.9}, need to be satisfied.  As a first step, one
easily sees that the product
 ${{\cal A}}_n{{\cal B}}_p$ must have the same sort of periodicity in $n$ as
does $c_n$, i.e.,
$c_{n+2} = -c_n$; therefore,
we need only resolve the products ${{\cal A}}_0{{\cal B}}_p$ and ${{\cal
A}}_1{{\cal B}}_p$.  The complete
set of requirements that this generates, on the (almost) arbitrary vectors
$\vec a_j$
and $\vec b_k$, are written out explicitly in the Appendix, in \Es{A11-14}.
\v
Once again, these
requirements have been found too complicated for us to completely resolve.
However, some useful comments may nonetheless be made.  For example, the
requirements include the statement that
$$\sum_{k=0}^\infty({{\cal A}}_0)^1{}_k(\vec b_m)^k \equiv ({{\cal A}}_0\,\vec
b_m)^1 = c_{2m+1}
= (-1)^m \ne 0\quad.\eqno(6.18)$$
\no From this we see that insisting that only
finitely many of the arbitrary (odd) columns are non-zero
is simply not allowed.  More precisely, if
we set $\vec b_m
\equiv 0$ for all $m$ greater than some integer, requirement \E{6.16} would
be impossible to satisfy.  By switching, and considering the column vector
$({{\cal B}}_0\vec
a_m)^1$, we can prove the same requirement for ${{\cal A}}_0$; therefore,
it is necessary that our matrices have infinitely many columns.  This lack of
finite-dimensional representations is already mirrored in the literature by
results of Shadwick, although he came to it from a rather different direction.
An alternative plausible mode for simplification would involve $\vec C_0$ more
intimately.  If one tries the ansatz that $\vec a_k = \alpha_k \vec C_0$, with
$\alpha_k$ a scalar, one easily finds contradictions to this
within the equations.
\v
Attempting, nonetheless, to find some route to follow further, we notice that
the requirements for, say ${\cal A}_0(\Lambda^T)^n\vec b_r$, in \Es{A11-A13} in
the Appendix,
insist that half of all the entries must explicitly be zero.  These entries
do suggest, {\bf and indeed allow}, yet an additional
specialization.  One may simply set to
zero {\bf all} the even-numbered elements of {\bf all} of our previously
arbitrary
vectors, which of course simplifies greatly the
equations, the general versions of which are listed
in the Appendix.  It is in fact this particular specialization that seems to
offer the intriguing connections with Toeplitz operators on Banach spaces.
Therefore,
we add this new assumption to the list, give new names to the resulting
arbitrary vectors that now determine our matrices, and display the resulting
equations in a compact form, involving (semi-infinite) Toeplitz matrices.
\v In this special case, where $(\vec a_j)^{2k} = 0 =
(\vec b_m)^{2n}$, we put an underline on the vectors to denote a symbol
that contains only the remaining elements:
$$(\ww{a}_j)^k \equiv (\vec a_j)^{2k+1}\qcomma (\ww{b}_m)^n \equiv (\vec
b_m)^{2n+1}\quad.\eqno(6.19)$$
\no The entire set of equations still to be satisfied are specified in
the Appendix, in component form, in \Es{A15-7}.
It is however much easier to write and
comprehend the equations when they are written in the form of Toeplitz
matrices.  For each $\ww{a}_j$ we define the matrices
$\bc{{\oua}}{j}$ and $\vec{\bc{{\sna}}{j}}$, where the last is conceived of as
a column vector with elements which are each row vectors:
$$\bc{{\oua}}{j}\equiv \sum_{k=0}^\infty (\ww{a}_j)^k\Lambda^k
= \pmatrix{\w{a}_j{}^0&\w{a}_j{}^1&\w{a}_j{}^2&\ldots\cr
0&\w{a}_j{}^0&\w{a}_j{}^1&\ldots\cr
0&0&\w{a}_j{}^0&\ldots\cr
0&0&0&\ldots\cr
\vdots&\vdots&\vdots&\ddots\cr},\;
\openup+2\jot\vec{\sna} \equiv \pmatrix{\ww{a}_0^{\;T}\cr\ww{a}_1^{\;T}\cr
\ww{a}_2^{\;T}\cr\ww{a}_3^{\;T}\cr\vdots\cr} =
\pmatrix{\w{a}_0{}^0&\w{a}_0{}^1&
\w{a}_0{}^2&\ldots\cr\w{a}_1{}^0&\w{a}_1{}^1&\w{a}_1{}^2&
\ldots\cr\w{a}_2{}^0&\w{a}_2{}^1&
\w{a}_2{}^2&\ldots\cr
\w{a}_3{}^0&\w{a}_3{}^1&\w{a}_3{}^2&\ldots\cr
\vdots&\vdots&\vdots&\ddots\cr
}\eqno(6.20a)$$
\no and identical objects made from the $\ww{b}_j$, and use
the binomial coefficients to define a matrix of the same type as
$\vec{\bc{{\sna}}{j}}$:
$$\openup+2\jot\eqalign{\vec{\bc{{\snm}}{j}} \equiv
\pmatrix{\vec{\fc{{\cal{M}}}{\ell}}{}_0{}^{\;T}\cr
\vec{\fc{{\cal{M}}}{\ell}}{}_1{}^{\;T}\cr
\vec{\fc{{\cal{M}}}{\ell}}{}_2{}^{\;T}\cr
\vdots\cr} & = \pmatrix{
\fc{{\cal{M}}}{\ell}{}_0{}^0&\fc{{\cal{M}}}{\ell}{}_0{}^1&
\fc{{\cal{M}}}{\ell}{}_0{}^2&\ldots\cr
\fc{{\cal{M}}}{\ell}{}_1{}^0&\fc{{\cal{M}}}{\ell}{}_1{}^1&
\fc{{\cal{M}}}{\ell}{}_1{}^2&\ldots\cr\fc{{\cal{M}}}{\ell}{}_2{}^0&
\fc{{\cal{M}}}{\ell}{}_2{}^1&
\fc{{\cal{M}}}{\ell}{}_2{}^2&\ldots\cr
\vdots&\vdots&\vdots&\ddots\cr}\quad,\cr
(\vec{\fc{{\cal{M}}}{\ell}}{}_m)^j & \equiv (-1)^{m+\ell+j}
\left\{{2m\choose 2j} -
{2m\choose 2j-2\ell -1}\right\}\quad.\cr}\eqno(6.20b)$$
\no Using these constructions, we may re-write the set of equations from
the Appendix in the very simple form
$$\bc{{\oua}}{j}\vec{\bc{{\snb}}{}} = \vec{\bc{{\snm}}{j}} =
\bc{{\oub}}{j}\vec{\bc{{\sna}}{}}\quad.\eqno(6.21)$$
\no Since every finite-dimensional sub-matrix of $\bc{{\oua}}{j}$ has a
straight-forwardly-defined determinant, one could suppose that it also has
one when the limit is taken, which would allow the formulation
$$ \vec{\bc{\sna}{}} = \big(\bc{{\oub}}{j}\big)^{\!-1}
\vec{\bc{{\snm}}{j}}\>,\quad
\forall j = 0,1,2,\ldots\quad.\eqno(6.22)$$
This gives a multitude of definitions of $\vec{\sna}$, all of which must of
course
hold simultaneously. In addition, the existence of an inverse for
$\vec{\bc{{\snm}}{0}}$ is
sufficient to allow a description of all the rest of the $\bc{\oub}{j}$ from
just the first one, $\bc{\oub}{0}$:
$$\bc{\oub}{j} = \vec{\bc{{\snm}}{j}}\big(\vec{\bc{{\snm}}{0}}\big)^{\!-1}
\bc{\oub}{0}\qquad,\qquad\bc{\oua}{j} =
\vec{\bc{{\snm}}{j}}\big(\vec{\bc{{\snm}}{0}}\big)^{\!-1}
\bc{\oua}{0}\quad.\eqno(6.23)$$
\v In order to justify the above manipulations with infinite-dimensional
matrices, we now show that all this may be put into the general theory
of Toeplitz operators acting on Banach spaces.\ft{50}
We attempt to give a brief introduction to these operators, but surely refer
interested parties to the literature mentioned in Ref. 50 for details,
justifications, and assorted special cases.  We describe these Toeplitz
operators by drawing the (usual) comparisons\ft{51} to simple multiplication
operators in, for instance, L${}^p$, as viewed via a point-wise definition
or their Fourier coefficients.
Let \setify{a,b,m,n,p,\, \ldots} be arbitrary functions on the circle, and
introduce
Fourier coefficients for each of them according to the following (usual)
pattern:
$$a(\theta) = \sum_{k=-\infty}^\infty\,a_k\,e^{ik\theta}
\qquad\Longleftrightarrow
\qquad a_k = {1\over
2\pi}\int_{-\pi}^{\pi}d\theta\,a(\theta)\,e^{-ik\theta}\quad.\eqno(6.24)$$
\no Thinking of these functions as defined in either space, their (pointwise)
product, as functions on the circle, takes quite a different form when
mapped into the space of Fourier coefficients.  We describe
these two operations as simply being two different representations of a {\it
multiplication operator}
associated with $a$:
$$(M_ab)(\theta) \equiv n(\theta) \equiv
a(\theta)b(\theta)\qquad\Longleftrightarrow
\qquad (M_ab)_k \equiv n_k  =
\sum_{n=-\infty}^\infty\,a_{k-n}\,b_n\quad.\eqno(6.25)$$
\no With this as background, the Toeplitz operator associated with the
function $a$, $T_a$, is created by this sort of a product, {\bf but} where
both $b$ {\it and} the product are restricted to have only non-negative Fourier
coefficients. (We say that they are elements of the Hardy space, ${\cl{H}}^+$,
the space of functions defined on the circle which have only non-negative
Fourier coefficients.)  The corresponding analogues to \Es{6.24-5} are now,
with $a,b\ \in\ {\cl{H}}^+$:
$$\eqalignno{b(\theta) = \sum_{k=0}^\infty\, b_k\,e^{ik\theta} \qquad
& \Longleftrightarrow
\qquad b_k = \cases{{1\over
2\pi}\int_{-\pi}^{\pi}d\theta\,b(\theta)\,e^{-ik\theta}&,\quad$k\ge 0$,\cr\cr
\qquad 0&,\quad$k\,<\,0$.\cr}&(6.26)\cr\cr
p(\theta) \equiv (T_a\,b)(\theta)\qquad & \Longleftrightarrow\qquad
p_k = \cases{\sum\limits_{n=0}^\infty\,a_{k-n}\,b_n\,&,\quad $k\ge 0$,\cr\cr
\qquad 0\,&,$\quad k < 0\ $.\cr}&(6.27)\cr}$$
\no For our purposes, we will also need related functions that lie within
the other Hardy space, ${\cl{H}}^-$,
the space of functions defined on the circle which have only non-positive
Fourier coefficients.  The (conjugate)\ft{51} function $\tilde a(\theta)$:
associated with $a(\theta)$, and the Toeplitz operator associated with $\tilde
a$,
are defined as follows:
$$\eqalignno{ a(\theta) = \sum_{k=-\infty}^{\infty}\,a_k\,e^{ik\theta}\qquad
& \Longleftrightarrow\qquad \tilde a(\theta) = \sum_{k=-\infty}^{\infty}\,
a_k\,e^{-ik\theta}\quad,&(6.28)\cr\cr
m(\theta) \equiv (T_{\tilde a}\,b)(\theta)\;,\qquad
& \Longleftrightarrow\qquad m_k  =
\cases{\sum\limits_{n=k}^\infty\,a_{n-k}\,b_n = \sum\limits_{p=0}^\infty
a_p\,b_{p+k}&,\quad $k\ge 0\ $,
\cr\cr \qquad 0 &,\quad $k < 0\ $.\cr}&(6.29)\cr}$$
\v The Toeplitz operator associated with a conjugate function, $T_{\tilde a}
\;:\;{\cl{H}}^-\times{\cl{H}}^+\longrightarrow{\cl{H}}^+$,
has exactly the same structure as the problem we want to solve.
In this language, we may re-write the equations from the Appendix
 in a mode that has somewhat
firmer foundation than the infinite matrices in \Es{6.21} and following.
In order to do this, we require the existence of five (related) sequences of
functions on the circle:
$$\{\aa^j,\,\pp^k,\, \mm^i\mid i,j,k = 0,\,1,\,\ldots\;\}\ \subseteq\ {\cl H}^+
\;,\quad\{\tilde{\aa}_\ell,\,\tilde{\pp}_m\mid \ell,m=0,\,1,\,\ldots\;\}\
\subseteq\ {\cl H}^-\quad.\eqno(6.30)$$
\no  Using these quantities, our constraints
take the following form, involving the Toeplitz action of
the conjugate functions on the other functions:
$$ \left(T_{{\tilde \pp}_\ell}\,\aa^j\right)(\theta) = {\mm}^j(\theta)
\equiv \sum_{m=0}^\infty\,{\mm}^j{}_m\,e^{im\theta} =
\left(T_{{\tilde \aa}_\ell}\,\pp^j\right)(\theta)\quad,\eqno(6.31)$$
\no The functions in question are related to our earlier ones, by
considering them as Fourier coefficients of
functions of two variables, $\aa(\theta,\zeta), \pp(\theta,\zeta)$, as follows:
$$\eqalign{\aa(\theta,\zeta)  =
&\sum_{\ell=0}^\infty\sum_{j=0}^\infty\,(\ww{a}_\ell)^j\,
e^{-ij\theta}\,e^{i\ell\zeta}\cr\cr
\sum_{s=0}^\infty\,(\ww{a}_\ell)^s\,e^{-is\theta}
= {\tilde \aa}_\ell(\theta)
= & \cases{{1\over 2\pi}\,\int_{-\pi}^{\pi}\,d\zeta \,\aa(\theta,\zeta) \,
e^{-i\ell\zeta} &,\quad $\ell \ge 0$,\cr\cr
\qquad 0&,\quad $\ell < 0$,\cr}\cr
\sum_{\ell=0}^\infty\,(\ww{a}_\ell)^j\,e^{i\ell\zeta}\phantom{-}
 = \aa^j(\zeta) = &
\cases{{1\over 2\pi}\,\int_{-\pi}^{\pi}\,d\theta\,\aa(\theta,\zeta)\,
e^{ij\theta}\phantom{-}&,\quad $j\ge 0$,\cr\cr\qquad 0 &, \quad$j <
0$,\cr}\cr}\eqno(6.32)$$
\no In terms of functions of one variable only, two pair of the functions
originate from the same pair of sources, our original set of matrix elements.
Therefore one may write a transformation between them:
$$\aa^j(\theta) = {1\over 2\pi}\sum_{\ell=0}^\infty\,e^{i\ell\theta}\,
\Four \!\!d\eta\,{\tilde \aa}_\ell(\eta)\,e^{ij\eta}\;,\hbox{~or~}\
{\tilde \aa}_\ell(\eta) = {1\over 2\pi}\sum_{n=0}^\infty\,e^{-in\eta}\,
\int_{-\pi}^\pi\!\!d\zeta\,\aa^n(\zeta)\,e^{-i\ell\zeta}\quad.\eqno(6.33)$$

\v With ${\mm}^j(\theta)$ being given, these are equations to determine the
unknown functions $\aa(\theta,\zeta)$ and $\pp(\theta,\zeta)$.  The sum
determining ${\mm}^j(\theta)$, from the coefficients given in \E{6.18c},
does not determine a function that exists in the usual Hilbert space;
nonetheless,
there are indeed Banach spaces in which it exists and makes good sense.
What is the optimum approach to considering these equations, and making sense
of
the set?  Unfortunately, we do not know, but do hope that some interest can be
generated among the considerable community of experts in the area of Toeplitz
problems.  We have considered an approach
via formal series, which may not be any more rigorous than our earlier
semi-infinite matrices; on the other hand, re-writing them in terms of integral
equations may well allow them to be re-considered from within some more
appropriate Banach space.  To do this, we begin with a {\it projection
operator} that projects general functions over the circle into ${\cl H}^+$:
$$b(\theta) =  {1\over 2\pi}\,\sum_{k=0}^\infty\,e^{ik\theta}
\int_{-\pi}^\pi d\phi\,b(\phi)\,e^{-ik\phi} = {1\over 2\pi}\int_{-\pi}^\pi
d\phi{b(\phi)\over 1-e^{i(\theta-\phi)}}\quad,\eqno(6.34)$$
\no which allows one to give an integral representation, valid at least in
terms of formal power series, of the Toeplitz operator:
$$ m(\theta) \equiv (T_{\tilde a}b)(\theta) =
 {1\over 2\pi}\int_{-\pi}^\pi d\phi {a(\phi)b(\phi)\over
1-e^{i(\theta-\phi)}}\quad,\eqno(6.35)$$
\no Using these projection operators, we may produce formal expressions that
re-write \Es{6.31} explicitly in terms of the 4 functions of one variable:
$${1\over 2\pi}\int_{-\pi}^\pi d\phi {\aa^j(\phi){\tilde \pp}_\ell(\phi)
\over 1-e^{i(\theta-\phi)}} = \mm^j(\theta) = {1\over 2\pi}\int_{-\pi}^\pi
d\phi\,{\pp^j(\phi){\tilde \aa}_\ell(\phi)\over 1-e^{i(\theta-\eta)}}\quad,
\eqno(6.36)$$
\no or a form with triple integrals can be written so that only the two
unknown functions of two variables need to be considered:
$$\eqalign{{1\over (2\pi)^3}\int_{-\pi}^\pi d\eta \int_{-\pi}^\pi
d\alpha & \int_{-\pi}^\pi d\beta\,{\pp(\alpha,\eta)\aa(\eta,\beta)\,
e^{-i(\ell\beta - j\alpha)}\over 1-e^{i(\theta-\eta)}}\quad =
\quad{\mm}^j(\theta)\cr
& =\quad {1\over (2\pi)^3}\int_{-\pi}^\pi d\eta \int_{-\pi}^\pi
d\alpha \int_{-\pi}^\pi d\beta\,{\pp(\eta,\beta)\aa(\alpha,\eta)\,
e^{-i(\ell\beta - j\alpha)}\over 1-e^{i(\theta-\eta)}}\cr}\eqno(6.37)$$
\no Multiplication of the above equation by $e^{-ij\eta}$ and summing on
all non-negative values of $j$ reduces the complexity of the equations, by
reducing the problem formulation to double integrals:
$$\int_{-\pi}^\pi d\phi\int_{-\pi}^\pi d\beta
{\pp(\eta,\phi)\aa(\phi,\beta) \,e^{-i\ell\beta}\over 1-e^{i(\theta-\phi)}}
= (2\pi)^2\mm(\eta,\theta) = \int_{-\pi}^\pi d\phi\int_{-\pi}^\pi d\beta
{\aa(\eta,\phi)\pp(\phi,\beta) \,e^{-i\ell\beta}\over 1-e^{i(\theta-\phi)}}
\eqno(6.38)$$
\v It is plausible that one or the other of the last two sets of equations,
taken {\it ab initio} would be a place from which the construction would
seem more reasonable and calculable.  From this point of view, of course,
one should consider how this would affect the original starting point,
which is simply an attempt to determine the completely general form of the
prolongation functions ${\bf F}$ and ${\bf G}$.  The
original indexed scalars, $A_{mn}^i$ were introduced because we needed to
express the commutator $[\b E_m,\b E_n]$.  Since we are now mapping the
Hardy space into itself, from the index (Fourier) approach to the
representation involving functions of $\theta$, this necessitates
the existence of (vector-field-valued)
functions on the circle, $\b E(\theta)$ and $\b J(\theta)$, such that
$$\b E_n = {1\over 2\pi}\Four \!\!d\theta\; e^{-in\theta}\b E(\theta)\quad
=\quad {-1\over 2\pi i}\oint{dt\over t}\;t^n\b E(\theta)
\quad,\eqno(6.39)$$
\no where the contour integral is taken around the unit circle in the
complex plane, along with a completely similar equation for $\b J_m$.
This alternative representation for $\b E_m$ also explains why
functions of two circle variables appeared in our equations,
since the two such variables correspond to the two
distinct occurrences of $\b E_n$ in the commutators under consideration.
Recalling the definitions of $\b F$ and $\b G$, from \Es{4.1-4},
this approach gives a complex-integral representation of them, involving
operators that might be from a Banach algebra, which is surely a
plausible alternative to a series expansion when that expansion is not
truncated:
$$\b F = {\h} {\bf R} -{1\over 2\pi i} \oint {dt\over t}\;\b E(t)
e^{{1\over 2} ut}\qcomma
\b G = -{\h}{\bf R} -{1\over 2\pi i}\oint{dt\over t}\;\b J(t)e^{-{1\over 2}
ut}\quad\eqno(6.40)$$
\vv
{\it Acknowledgments:}  We would like to express our appreciation
for the very considerable assistance with symbolic-calculation
computer codes given to us by Michael Wester.  Our understanding of
gauge transformations was greatly increased by discussions with Mikhail
Saveliev.  As well one of us
(JDF) has benefited from
extensive discussions with Mark Hickman, Cornelius Hoenselaers and
Alice Fialowski concerning these problems.
\veject

\v\v\no{\bf Appendix:~~Some Technical Details of the Resolution of $\qb_2$}\v
\no\underbar{Requirement One}:\v
The skew-symmetry condition, \Es{6.8}, and the recursion relation, \Es{6.9},
impose non-trivial constraints on the semi-infinite matrices, ${{\cal A}}_0$.
To determine a form that satisfies these constraints, we begin with an
important part of the skew symmetry constraint, that the diagonal
elements must vanish:
$({{\cal A}}_0)^k{}_0 = 0, ({{\cal A}}_1)^k{}_1 = 0,\;\ldots\,,\; ({{\cal
A}}_i)^k{}_i = 0$.  When
these simple statements are mixed with the recursion relation for the full
matrices, we obtain series that allow us to determine other elements of ${{\cal
A}}_0$.
The first of these statements simply says that the first column of
the matrix is zero.  The lowest non-trivial examples are
$$\eqalign{0 = ({{\cal A}}_1)^k{}_1  = -({{\cal A}}_0)^{k-1}_{\quad 1} -
({{\cal A}}_0)^k{}_2\quad,\
& \Longrightarrow ({{\cal A}}_0)^k{}_2 = -({{\cal A}}_0)^{k-1}_{\quad
1}\quad,\cr
0 = ({{\cal A}}_2)^k{}_2 = ({{\cal A}}_0)^{k-2}_{\quad2} + 2({{\cal
A}}_0)^{k-1}_{\quad3}+({{\cal A}}_0)^k{}_4\quad,\
& \Longrightarrow ({{\cal A}}_0)^k{}_4 = -2({{\cal
A}}_0)^{k-1}_{\quad3}-({{\cal A}}_0)^{k-2}_{\quad2}\quad.
\cr}\eqno(A1)$$
\no The general term of this relationship takes the form
$$\eqalign{ 0 = \left({{\cal A}}_i\right)^k{}_i & =
(-1)^i\sum_{m=0}^i{i\choose m}\left({{\cal A}}_0\right)^{k-i+m}_{\qquad i+m}\cr
\Longrightarrow \qquad \left({{\cal A}}_0\right)^k{}_{2i} &  =
-\sum_{m=0}^{i-1}{i\choose m}\left({{\cal A}}_0\right)^{k-i+m}_{\qquad i+m} =
-\sum_{n=1}^i{i\choose n}\left({{\cal A}}_0\right)^{k-m}_{\qquad
2i-n}\quad.}\eqno(A2)$$
\no  By looking in detail at the procedure in \Es{A1},
one can use the equation for $i=1$ to determine $({{\cal A}}_0)^k{}_2$,
then set $i=2$ and
insert into it the just-determined value for $({{\cal A}}_0)^k{}_2$ so that
$({{\cal A}}_0)^k{}_4$ can be determined.  After that, both values are inserted
into
the $i=3$ equation to determine $({{\cal A}}_0)^k{}_6$.  Therefore, operating
in this
doubly-recursive manner, we may sequentially
solve an $i$-term recursion relation for
each element of the $2i$-th column.  The result is given in \Es{6.14}, where
the process described below determines the values of the coefficients,
generated by sums of products of binomial coefficients, needed for this.
\v To display the solutions
for arbitrary values of $i$, we first divide up the sum, in \Es{A2}, into those
terms involving elements from {\it even} columns and those
involving elements from {\it odd} columns:
$$\left({{\cal A}}_0\right)^k{}_{2i}  = -\sum_{p=0}^{[{i-1\over 2}]}{i\choose
2p+1}
\left({{\cal A}}_0\right)^{k-2p-1}_{\qquad 2i-2p-1} \quad - \quad
\sum_{q=1}^{[{i\over 2}]}
{i\choose 2q}\left({{\cal A}}_0\right)^{k-2q}_{\qquad 2i-2q}\quad.\eqno(A3)$$
\no   Within the expansion of
some arbitrary even column, the highest-appearing odd column occurs only once,
with coefficient ${i\choose 1}$; however, the next-highest odd column occurs
twice, so that its contribution to the final sum must be determined by summing
those two coefficients.  The next columns appears three times, etc.
We must re-sum the series so as to determine the coefficient
that multiplies entries from $({{\cal A}}_0)_{2j-2r}$ in the expansion to
determine
$({{\cal A}}_0)_{2j}$.  These coefficients we denote by $\{\cq_{j,r}\mid
j=1,2,3,\ldots\>;\>r=0,1,2,\ldots,r-1\}$, and specify the following
 recursive method of obtaining them:
$$\displaylines{\hfill\cq_{j,0} = 1\>,\qquad \cq_{j,r} =
-\sum_{k=0}^{r-1}{j-k\choose 2r-2k}\,\cq_{j,k}\;,\quad
 r = 0,1,\>\ldots\>r-1\quad,\hfill\llap{(A4)}\cr
\cq_{1,r}\;\colon\; [1]\quad,\cr
\cq_{2,r}\;\colon\; [1, -1]\quad,\cr
\cq_{3,r}\;\colon\; [1, -3, 3]\quad,\cr
\cq_{4,r}\;\colon\; [1, -6, 17, -17]\quad,\cr
\cq_{5,r}\;\colon\; [1, -10, 55, -155, 155]\quad,\cr
\cq_{6,r}\;\colon\; [1, -15, 135, -736, 2073, -2073]\quad.\cr
\ldots\cr}$$
\no Having those coefficients for each $({{\cal A}}_0)_{2j-2r}$, we may now
sum these quantities multiplied by each appropriate weight factor, to determine
the coefficients actually desired, which we will refer to by
the notation $\{\cp_{i, n}\mid i = 1, 2, \ldots\>;\>
n = 0, 1,\>\ldots\>i-1\}$:
$$\displaylines{\hfill\cp_{i,w} \equiv  -\sum_{r=0}^w{i-r\choose 2w+1-2r}
\,\cq_{i,r}\quad,\hfill\llap{(A5)}\cr
\cp_{1,r}\;\colon\; [-1]\quad,\cr
\cp_{2,r}\;\colon\; [-2, 1]\quad,\cr
\cp_{3,r}\;\colon\; [-3, 5, -3]\quad,\cr
\cp_{4,r}\;\colon\; [-4, 14, -28, 17]\quad,\cr
\cp_{5,r}\;\colon\; [-5, 30, -126, 255, -155]\quad,\cr
\ldots\cr}$$
\no and we always take the binomial coefficients as zero when negative
factorials appear in the denominator.
To display more clearly the general form of a ${{\cal A}}_0$ that satisfies the
first 2 of our requirements, we write out the
elements nearest the upper left corner:\v
\moveleft10pt\vbox{$$\pmatrix{0&(a_0)^0&0&(a_1)^0&0&(a_2)^0&0&\cdots\cr
0&(a_0)^1&-(a_0)^0&(a_1)^1&-2(a_1)^0&(a_2)^1&-3(a_2)^0&\cdots\cr
0&(a_0)^2&-(a_0)^1&(a_1)^2&-2(a_1)^1&(a_2)^2&-3(a_2)^1&\cdots\cr
0&(a_0)^3&-(a_0)^2&(a_1)^3&-2(a_1)^2+(a_0)^0&(a_2)^3&-3(a_2)^2+5(a_1)^0&\cdots\cr
0&(a_0)^4&-(a_0)^3&(a_1)^4&-2(a_1)^3+(a_0)^1&(a_2)^4&-3(a_2)^3+5(a_1)^1&\cdots\cr
0&(a_0)^5&-(a_0)^4&(a_1)^5&-2(a_1)^4+(a_0)^2&(a_2)^5&-3(a_2)^4+5(a_1)^2-3(a_0)^0&\cdots\cr
0&\vdots&\vdots&\vdots&\vdots&\vdots&\vdots&\ddots\cr}$$}\vskip-40pt
$${}\eqno(A6)$$
\no Of course there is an identical expression for ${{\cal B}}_0$, made from
arbitrary
vectors $\vec b_k$.
\v\no\underbar{Requirement Two}:\v
The second set of requirements, imposed by the eigenvalue-type equations,
\Es{6.10}, may also be explicitly resolved.  Both sides of the equations
involve an infinite sum of elements of ${{\cal A}}_i$ with
elements of $\vec C_k$.  However, since the $c_{j+k} = (\vec C_k)^j$ take on
only the values $0, 1, 0 , -1$, and then repeat, such sums only generate
alternating-sign sums of alternating elements of the matrices ${{\cal A}}_i$.
Defining (subscripted) vectors $\vec u_{ji}$,
$$(\vec u_{ji})_\ell\ \equiv\ \cases{(-1)^{j/2}\sum_{m=0}^\infty(-1)^m({{\cal
A}}_i)^{2m+1}{}_\ell,& $j$ even,\cr \cr
(-1)^{(j-1)/2}\sum_{m=0}^\infty(-1)^m({{\cal A}}_i)^{2m}{}_\ell, &$j$
odd,\cr}\eqno(A7)$$
 allows us to re-write \Es{6.10} and display the periodicity they possess:
$$ (\vec u_{ji})_\ell - (\vec u_{ij})_\ell -
(\vec u_{\ell i})_j = 0\quad.\eqno(A8)$$
\no The general recursion relation for these matrices allows to restrict our
attention to ${{\cal A}}_0$.  This resolution does of course involve entries
from both
the (arbitrary) odd columns, and the even columns, which are given
already as sums of the odd columns, via \Es{6.14}.  Therefore the
results will involve the coefficients $\cp_{i,r}$ just determined, at
\Es{A6}.
\v We give explicit names to these sums for the matrix ${{\cal A}}_0$:
$$ \eqalign{R_{j}\equiv\sum_{m=0}^\infty (-1)^{j+m}(\vec a_j)^{2m} = \vec
C_1\cdot\vec a_j & \quad,\quad
S_{j}\equiv\sum_{m=0}^\infty (-1)^{j+m}(\vec a_j)^{2m+1}=
\vec C_0\cdot\vec a_j\quad.\quad \cr}\eqno(A9)$$
\no When the requirements are written in terms of these quantities,
they induce recursion relations for them.  For example, values of
\setify{0,1,2}
for the indices gives us $3S_0 + S_1 = 0$.  As the indices grow, the number of
terms in the relations grows.
After some little algebra, we acquire the relations in  the form of
$(n+2)$-term
recursion relations for the desired quantities, $R_n$ and $S_n$:
$$(n-1)R_n =  - R_0 + \sum_{h=1}^n\cp_{n+1,h}R_{n-h}\quad,\quad
S_n  = S_0 - 2\sum_{h=1}^{n}\cp_{n,h-1}S_{n-h}\quad,\eqno(A10)$$
\no along with identical relations for $U_n$ and $V_n$, made analogously with
the
sums made from the elements of ${{\cal B}}_0$.
The equations require an initial input of values for $R_0$,
$R_1$, and $S_0$, but then allow explicit, recursive calculation of the values
of
all the others.  This calculation results in the numerical sequences, for
$w_m$ and $q_j$, given in \Es{6.16}.
\v\v
\no\underbar{Requirement Three}:\v
Taking the form already determined for each of
the matrices ${{\cal A}}_n$ and ${{\cal B}}_m$,
the quadratic requirements may be written in terms of
the various quantities ${{\cal A}}_0(\Lambda^T)^{n}\vec b_j$ and
${{\cal B}}_0(\Lambda^T)^{n}\vec a_j$:
$$ {{\cal A}}_0(\Lambda^T)^n\vec b_r\quad = \quad\vec\ck^{\,n}{}_\ell
\quad = \quad{{\cal B}}_0(\Lambda^T)^n\vec a_r\quad,
\eqno(A11)$$
\no where the 2-index vectors $\vec\ck^{\,n}{}_\ell$ have elements that are
always
integers.  The equations may be written out explicitly in terms of the vectors
themselves, or
in terms of the components, with the latter actually being
 simpler-appearing set of expressions.  The numerical quantities have the form
$$\displaylines{\hfill\vec\ck^{\,n}{}_\ell \equiv
\sum_{p=0}^{\left[n/2\right]}(-1)^{n+p+\ell}{n\choose 2p}
\left(\Lambda^T\right)^{n-2p}\vec\ell - \sum_{q=0}^{\left[(n-1)/2\right]}
(-1)^{n+q}{n\choose 2q+1}\left(\Lambda^T\right)^{n+2\ell-2q}\vec\ell\qquad
\hfill\llap{(A12)}\cr
\hfill\left(\vec\ck^{\,n}{}_\ell\right)^i =
\cases{(-1)^{m+r+j}\left({2m+1\choose
2j-1} - {2m+1\choose 2(j-r-1)}\right),& for even $i \equiv 2j$
and odd $n\equiv 2m+1$,\cr
0,& for odd $i$ and odd $n$\cr
0,& for even $i$ and even $n$\cr
(-1)^{m+r+j}\left({2m\choose 2j} - {2m\choose 2(j-r)-1}\right)\>,& for odd $i
\equiv 2j+1$ and even $n\equiv 2m$,\cr}\hfill\llap{(A13)}\cr}$$
\no where the vector, $\vec \ell$ is just $(0,1,0,0,0,\ldots)^T$, the symbol
$\left[n/2\right]$ means the greatest integer in $n/2$. The left- and
right-hand
sides of the equations must also be decomposed:
$$  {{\cal A}}_0(\Lambda^T)^n\vec b_r =
\cases{\sum\limits_{k=0}^\infty\left\{(\vec b_\ell)^{2k}
\vec a_{m+k} + \sum\limits_{u=0}^{m+k}\cp_{m+k+1,u}(\vec
b_\ell)^{2k+1}(\Lambda^T)^{2u+1}
\vec a_{m+k-u}\right\}\>,&\cr \hfil\qquad\qquad\hbox{for odd}\ n\equiv 2m+1,
&\cr\cr
\sum\limits_{k=0}^\infty\left\{(\vec b_\ell)^{2k+1}
\vec a_{m+k} + \sum\limits_{u=0}^{m+k-1}\cp_{m+k,u}(\vec
b_\ell)^{2k}(\Lambda^T)^{2u+1}
\vec a_{m+k-1-u}\right\}\>,& \cr \hfil\qquad \qquad\hbox{for even}\ n\equiv 2m,
&\cr}\eqno(A14)$$
\no and a similar expression for
${{\cal B}}_0(\Lambda^T)^n\vec a_r$, with the roles of $\vec
b_j$ and $\vec a_i$ reversed.  When inserted into \Es{A11}, these constitute
the most general form of the quadratic conditions.
\v
As discussed in Section VI, the equations just displayed seem to be too
difficult to fully digest.  On the other hand, as noted there, the fact that
half of the quantities in \Es{A13} are in fact simply zero causes one to think
of proposing that a simpler solution might still be obtained in half of the
elements of our arbitrary vectors were set directly to zero.  When this is
done, the remaining
half of the equations, still to be satisfied, take the following form:
$$\eqalignno{&\sum_{k=0}^\infty (\ww{b}_\ell)^k\ww{a}_{m+k}  =
\vec{\fc{{\cal{M}}}{\ell}}{}_m
 = \sum_{k=0}^\infty (\ww{a}_\ell)^k\ww{b}_{m+k} \quad,&(A15)\cr
\sum_{k=0}^\infty\sum_{u=0}^{m+k}\cp_{m+k+1,u}(\ww{b}_\ell)^k
&(\ww{a}_{k+m-u})^{j-u-1}
 = (-1)^{m+\ell+j}\left\{{2m+1\choose 2j-1} - {2m+1\choose 2j -2l
-2}\right\}\cr
& = \sum_{k=0}^\infty\sum_{u=0}^{m+k}\cp_{m+k+1,u}(\ww{a}_\ell)^k
(\ww{b}_{k+m-u})^{j-u-1}\quad,&(A16)\cr}$$
\no where the elements of the vectors $ \vec{\fc{{\cal{M}}}{\ell}}{}_m$ are
simply combinations
of binomial coefficients:
$$ (\vec{\fc{{\cal{M}}}{\ell}}{}_m)^j  \equiv (-1)^{m+\ell+j}
\left\{{2m\choose 2j} -
{2m\choose 2j-2\ell -1}\right\}\quad.\eqno(A17)$$

\veject
\centerline{REFERENCES}
\parindent=0pt
\baselineskip=14pt
\def\hi{\hangindent=20truept}
\def\bn#1{{\bf #1}}
\def\IT#1{{\it #1 \/}}
\frenchspacing

\hi ~1.  Very early references of which we are aware are the following:
S. Lie,  ``Z\"ur Theorie der Fl\"achen Konstanter Kr\"ummung,"
Arch. Math. Naturvidensk \b 5, III. 282-306 (1880)
and IV. 328-358 (1880).  Work of
P. L. Chebyshev,  ``On the cutting of clothes," apparently presented to
the French Academy of Sciences in 1878; a Russian translation of some material,
kept in the archives of the Akademiei Nauk, has been published in Usp. Mat.
Nauk {\bf 1}, 38-42 (1946).  We learned this from an article of Poznyak, who
refers to our equation as the Chebyshev equation:  E. G. Poznyak
``Geometric Investigations connected with $z_{xy} =
\sin z$," J. Soviet Math. \b{13}, 677-686 (1980), from Itogi Nauki i Tekhniki.
Problemy Geometrii, \b 8, 225-241 (1977).  A. V. B\"acklund,
``Einiges uber Curaven und Fl\"achentransformationen,'' Lund Universitets
Arsskrift \b{10}, (1875), ``Theorie der Fl\"achentranformationen,'' Math. Ann.
{\bf 19}, 387-422 (1882), and ``Om ytor med konstant negativ kr\"okning,"
Lunds Universitets Arsskrift Avd. \b{19}, (1883).

\hi ~2.  C. Rogers and W. F. Shadwick, {\it B\"acklund Transformations and
Their Applications\/} (Academic Press, New York, 1982).

\hi ~3.  A. C. Scott, F. Y. F. Chu, and D. W. McLaughlin, ``The soliton:  a new
concept in applied science," Proc. IEEE \b{61}, 1443-1483 (1973).

\hi ~4.  A.V. Zhiber and A.B. Shabat, ``Klein-Gordon equations with a
nontrivial
group," Sov. Phys. Dokl. \b{24}, 607-9 (1979), and
``Equations of Liouville type," Sov. Math. Dokl.
\b{20}, 1183-7 (1979).  

\hi ~5. H.-H. Chen, ``General Derivation of B\"acklund Transformations from
Inverse Scattering Problems," Phys. Rev. Lett. \b{33}, 925-8 (1974).

\hi ~6.  S. G. Byrnes, ``B\"acklund transformations and the equation $z_{xy}
= F(x,y,z)$,'' J. Math. Phys. {\bf 17}, 836-842 (1976).  

\hi ~7.  A. N. Leznov and M. V. Saveliev, ``Representation of
Zero Curvature for the
System of Non-linear Partial Differential Equations $x_{\alpha,z\overline z} =
exp(kx)_\alpha$ and its Integrability," Lett. Math. Phys. \b 3, 489-494 (1979).
A. N. Leznov, V. G. Smirnov, and A. B. Shabat, ``The Group of Internal
Symmetries and the Conditions of Integrability of Two-Dimensional Dynamical
Systems," Theor. \& Math. Phys. \b{51}, 322-330 (1982).

\hi ~8.  R.K. Dodd and R. K. Bullough, ``Polynomial conserved densities for
the sine-Gordon equations," Proc. R. Soc. Lond. A. \bn{352}, 481-503 (1977);
Allan P. Fordy and John Gibbons, ``Integrable Nonlinear Klein-Gordon
Equations and Toda Lattices," Commun. Math. Phys. \bn{77}, 21-30 (1980).

\hi ~9.  Shiing-Shen Chern and Chuu-Lian Terng, ``An Analogue of B\"acklund's
Theorem in Affine Geometry," Rocky Mountain Journal of Mathematics \b{10},
105-124 (1980).

\hi 10.  G. L. Lamb, Jr., ``B\"acklund transformations for certain nonlinear
evolution equations,'' J. Math. Phys. \b{15}, 2157-65 (1974), and ``Propagation
of Ultrashort Optical Pulses," Phys. Lett. A\b{25}, 181-2 (1967) are two
appropriate places to look for historical details.

\hi 11.  Hugo Wahlquist and Frank Estabrook,  ``B\"acklund Transformation for
Solutions of the Korteweg-de Vries Equation,'' Phys. Rev. Lett. \b{31}, 1386-90
(1973); Frank Estabrook and Hugo Wahlquist, ``Prolongation structures of
nonlinear evolution equations,'' J. Math. Phys. {\bf 16}, 1-7 (1975), and
Hugo Wahlquist and Frank Estabrook, ``Prolongation structures of nonlinear
evolution equations.  II.'' J. Math. Phys. {\bf 17}, 1293-7 (1976).

\hi 12.  Elie Cartan, {\it Les syst\`emes diff\'erentiels ext\'erieurs et
leurs applications g\'eom\'etriques} (Hermann, Paris, 1945).

\hi 13.  F. Pirani, D. Robinson and W. Shadwick, {\it Local Jet Bundle
Formulation
of B\"acklund Transformations\/}, Mathematical Physics Studies, Vol. 1 (Reidel,
Dordrecht, 1979).

\hi 14.  B. Kent Harrison, ``Unification of Ernst-equation B\"acklund
transformations using a modified Wahlquist-Estabrook technique," J. Math.
Phys. \b{24}, 2178-87 (1983).

\hi 15.  P.H.M. Kersten, {\it Infinitesimal symmetries:  a computational
approach} (Centrum voor Wis\-kunde en Informatica, Centre for Mathematics
and Computer Science, P.O. Box 4079, 1009 AB Amsterdam, The Netherlands, 1987).
Earlier references include P.K.H. Gragert, P.H.M. Kersten, and R. Martini,
``Symbolic Computations in Applied Differential Geometry,"
Acta Applicandae Mathematicae \b{1}, 43-77 (1983).

\hi 16.  James P. Corones and Frank J. Testa, ``Pseudopotentials and Their
Applications,"  in {\it B\"acklund Transformations} Lecture Notes in
Mathematics,
Vol. 515 (Springer-Verlag, New York, 1976); James P. Corones,
``Solitons, pseudopotentials, and certain Lie algebras," J. Math. Phys.
\bn{18}, 163-4 (1977);  E.N. Glass \& D.C. Robinson,
``A nilpotent prolongation of the Robinson-Trautman equation,"
J. Math. Phys. \bn{25}, 3342-6 (1984).

\hi 17.  Wolfgang Karl Schief, Diplomarbeit, ``Zur Prolongationstheorie,'' Max
Planck Institut f\"ur Astrophysik, Garching, 1989 (unpublished); Cornelius
Hoenselaers and Wolfgang K. Schief,
``Prolongation structures for Harry Dym type equations and B\"acklund
transformations of CC ideals," J. Phys. \bn{A25}, 601-22 (1992).

\hi 18.  H. C. Morris, ``B\"acklund Transformations and the sine-Gordon
Equation,"
in {\it The 1976 Ames Research Center (NASA) Conference on the Geometric Theory
of Non-Linear Waves}, edited by R. Hermann (Math. Sci. Press, Brookline, Mass.,
1977).

\hi 19.  Minoru Omote, ``Prolongation structures of nonlinear equations and
infinite-dimensional algebras," J. Math. Phys. \b{27}, 2853-60 (1986).

\hi 20.  C. Hoenselaers,
``The Sine-Gordon Prolongation Algebra,"
Progress of Theor. Physics \b{74}, 645-654 (1985).

\hi 21.  W.F. Shadwick,
``The B\"acklund problem for the equation $d^2z/dx^1dx^2 = f(z)$,"
J. Math. Phys. \bn{19}, 2312-2317 (1978), is an important
reference for our purposes.

\hi 22. R.K. Dodd and J. D. Gibbon, ``The prolongation structures
of a class of nonlinear evolution equations," Proc. R. Soc. Lond.
A. \b{359}, 411-433 (1978).

\hi 23. J. D. Finley, III and John K. McIver,
``Prolongations to Higher Jets of Estabrook-Wahlquist Coverings for PDE's,"
Acta Applicandae Mathematicae \bn{32}, 197-225 (1993).

\hi 24.  C. Hoenselaers, ``More Prolongation Structures,"
Progress of Theor. Physics \b{75}, 1014-29 (1986),
and ``Equations admitting O(2,1)$\times$R($t,t^{-1})$ as a prolongation
algebra,"
J. Phys. A \b{21}, 17-31 (1988).

\hi 25. H.N. van Eck,
``The explicit form of the Lie algebra of Wahlquist and Estabrook.  A
presentation
problem." Proc. Kon. Ned. Akad. Wetensch, Ser. A {\bf 86}, 149-164 and
165-172  (1983).

\hi 26.  J.H.B. Nijhof and G.H.M. Roelofs,
``Prolongation structures of a higher-order nonlinear Schr\"odinger equation,"
J. Phys. \bn{A25}, 2403-16 (1992),
 G. H. M. Roelofs and R. Martini,
``Prolongation structure of the KdV equation in the bilinear form of Hirota,"
J. Phys. \bn{A23}, 1877-84 (1990),
W. M. Sluis and P.H.M. Kersten,
``Non-local higher-order symmetries for the Federbush model,"
J. Phys. \bn{A23}, 2195-2204 (1990).

\hi 27.  W.F. Shadwick,
``The KdV prolongation algebra" J. Math. Phys. \bn{21}, 454-61 (1980).

\hi 28.  Frank B. Estabrook, ``Differential Geometry Techniques for Sets of
Nonlinear
Partial Differential Equations," in {\it Partially Integrable Evolution
Equations in Physics,} edited by R. Conte and N. Boccara (Kluwer Academic
Pubs.,
Netherlands, 1990), p. 413-434.

\hi 29.  Theo van Bemmelen, ``Applications of coverings and non-local
symmetries,'' J. Phys. A.{\bf 26}, 6409-6420 (1993).  

\hi 30.  A.M. Vinogradov,
``Local Symmetries and Conservation Laws,"
Acta Applicandae Mathematicae {\bf 2},
21-78 (1984), and ``Symmetries and Conservation Laws of Partial Differential
Equations:  Basic Notions and Results,"
Acta Applicandae Mathematicae {\bf 15}, 3-21 (1989).
Although he has written many articles on these subjects,
these particular ones give a good
introduction and review of the material.  There are also two recent textbooks
written by him and co-authors:  D.V. Alekseevskii,  V.V. Lychagin and A.M.
Vinogradov, {\it Category of nonlinear
differential equations} (Gordon and Breach, New York, 1986) and {\it Geometry
I} (Springer-Verlag, Berlin, 1991).

\hi 31.  I. S. Krasil'shchik and A. M. Vinogradov,
``Nonlocal Symmetries and the Theory of Coverings:  An Addendum to A. M.
Vinogradov's Local Symmetries and Conservation Laws," Acta
Applicandae Mathematicae {\bf 2},  79-96 (1984).

\hi 32.  I. S. Krasil'shchik and A.M. Vinogradov,
``Nonlocal Trends in the Geometry of Differential Equations:
Symmetries, Conservation Laws, and B\"acklund Transformations," Acta
Applicandae Mathematicae {\bf 15}, 161-209 (1989).

\hi 33.  V.E. Zakharov and A.B. Shabat,
``Exact Theory of Two-Dimensional Self-Focusing and One-Dimensional
Self-Modulation of Waves in Nonlinear Media,"
J.E.T.P. \bn{34}, 62-9 (1972),
``A Scheme for Integrating the Nonlinear Equations of Mathematical Physics
by the Method of the Inverse Scattering Problem. I" Functional
Analysis \& Applications
\bn8, 43-53 (1974), and
``Integration of Nonlinear Equations of Mathematical Physics by the Method
of Inverse Scattering. II" Functional Analysis \& Applications
\bn{13} 13-22 (1976).

\hi 34.  A. N. Leznov, M. V. Saveliev, {\it Group-Theoretical Methods for
Integration of Nonlinear Dynamical Systems} (Birkh\"auser Verlag, Basel, 1992).

\hi 35.  B.G. Konopelchenko, W. Schief, C. Rogers,
``A (2+1)-dimensional sine-Gordon system:  its auto-B\"acklund transformation,"
Physics Letters {\bf A 172}, 39-48 (1992); Yu-kun Zheng and W. L. Chan,
``Gauge and B\"acklund transformations for the variable coefficient
higher-order modified Korteweg-de Vries equation,"
J. Math. Phys. {\bf 29}, 2570-5 (1988).
Yi Chen, Walter Strampp, and You-Jin Zhang, ``Bilinear B\"acklund
transformations
for the KP and k-constrained KP hierarchy," Phys. Letters A{\bf 182}, 71-76
(1993);
Lin-Lie Chau, J. C. Shaw, and H. C. Yen, ``Solving the KP Hierarchy by Gauge
Transformations," Commun. Math. Phys. {\bf 149}, 263-278 (1992).

\hi 36.  Karen Uhlenbeck, ``On the connection between harmonic maps and
the self-dual Yang-Mills and the sine-Gordon equations," J. Geometry and
Physics {\bf 8}, 283-316 (1992);  C. Sophocleous and J. G. Kingston,
``A class of B\"acklund transformations for equations of the type
$u_{xy} = f(u,u_x)$,'' J. Math. Phys. {\bf 32}, 3176-83 (1991); R. Beutler,
``Positon solutions of the sine-Gordon equation,'' J. Math. Phys. {\bf 34},
3098-3109 (1993).

\hi 37.  J. D. Finley, III, ``The Robinson-Trautman Type III Prolongation
Structure
Contains $K_2$,'' to be published.

\hi 38.  Private communication from Mark Hickman.

\hi 39.   B. Kent Harrison and Frank B. Estabrook,
``Geometric Approach to Invariance Groups and Solution of Partial
Differential Systems," J. Math. Phys. \bn{12},
653-65 (1971);  Peter J. Olver, {\it Applications of Lie Groups to Differential
Equations} (Springer-Verlag, New York, 1986).

\hi 40.  See p. 114 ff of Ref. 13 for a description of effective subideals, and
some examples for this equation.

\hi 41.  P. Molino, J. Math. Phys. \b{25}, 2222-5 (1984), discusses why this
can be done for the KdV equation in a way which easily generalizes to the case
when the independent variables do not appear explicitly in the equation, and
that equation is quasilinear.

\hi 42.  Mark Hickman and JDF spent a few weeks attempting to find general
integrals for this problem.  Krasil'shchik, in Ref. 32, gives only such
one-dimensional coverings.

\hi 43.  Frank B. Estabrook, ``Moving frames and prolongation algebras,"
J. Math. Phys. \b{23}, 2071-6 (1982).

\hi 44.  Shlomo Sternberg, {\it Lectures on Differential Geometry\/} (Chelsea
Publ. Co., New York, 1983).  See Ch. 5. Our form is simply a coordinate
presentation of the action on a Lie algebra of
the flow of a vector field generated by the adjoint action of another element
of the Lie algebra.

\hi 45.  R. Hermann, {\it Differential Geometry and the Calculus of
Variations},
2nd Ed., Interdisciplinary Mathematics XVII (Math. Sci. Press, Brookline,
Mass., 1977).
See Ch. 18, Accessibility Problems for Path Systems.

\hi 46.  In Ref. 34, \S3.1.2, those authors begin from a set of potential
equations which has the sine-Gordon
equation as an integrability condition.  This allows them to generate
 the center of $A_1^{(1)}$ itself, which distinguishes it from
$A_1\otimes\complx[\lambda^{-1},\lambda]$.  Our approach seems unable to
do this.

\hi 47.  Although re-discovered by Dodd and Bullough, in Ref. 8,
this equation was first studied by M. Tzitz\'eica,
``Sur une nouvelle classe de surfaces," Comptes Rendu Acad.
Sci. {\bf 150}, 955-6 (1910), as I learned from Graeme A. Guthrie.

\hi 48.  L. O'Raifeartaigh, {\it Group structure of gauge theories} (Cambridge
Univ. Press, Cambridge, UK, 1986).

\hi 49.  Cornelius Hoenselaers, private communication.

\hi 50.  Ulf Grenander and Gabor Szeg\"o, {\it Toeplitz Forms and Their
Applications} (University of California Press, Berkeley, Calif., 1958),
describes
earlier work in this area.
R. G. Douglas, \IT{Banach algebra techniques in the theory of
Toeplitz operators,} Conference Board of the Mathematical
Sciences, No. 15 (American Mathematical Society, Providence, RI, 1973), and
 Ref. 51 are modern presentations of what seems to us to be near the
``state of the art.''

\hi 51.  Albrecht B\"ottcher and Bernd Silbermann, \IT{Analysis of Toeplitz
Operators} (Springer-Verlag, Berlin, 1990).  See p. 57 for this particular
notion of
conjugate functions.

\vfill
\eject
\bye